\documentclass[aps,prl,showpacs,amsmath,amssymb,sort&compress,comma,nofootinbib,floatfix,twocolumn,superscriptaddress]{revtex4-1}
\usepackage{graphicx}
\usepackage{subfigure}  
\usepackage{bm}
\usepackage{times}
\usepackage{graphicx}
\usepackage[colorlinks,citecolor=blue,linkcolor=blue]{hyperref}
\usepackage{array}
\usepackage{float}
\usepackage{colortbl}
\usepackage{multirow}
\usepackage{tabularx}
\begin{document}

\newcommand{\etal}{{\it et al. }\/}
\newcommand{\gtwid}{\mathrel{\raise.3ex\hbox{$>$\kern-.75em\lower1ex\hbox{$\sim$}}}}
\newcommand{\ltwid}{\mathrel{\raise.3ex\hbox{$<$\kern-.75em\lower1ex\hbox{$\sim$}}}}

\title{Tunneling Magnetoresistance in Junctions Composed of
Ferromagnets and Time-Reversal Invariant Topological Superconductors}

\author{Zhongbo Yan}
\email[]{yzhbo@mail.tsinghua.edu.cn}
\affiliation{Institute for Advanced Study, Tsinghua University,
Beijing, 100084, China}
\affiliation{Department of Modern Physics, University of Science and
Technology of China, \\
Hefei, Anhui Province, 230026, China}

\author{Shaolong Wan}
\affiliation{Department of Modern Physics, University of Science and
Technology of China, \\
Hefei, Anhui Province, 230026, China}

\begin{abstract}
Tunneling Magnetoresistance between two ferrromagnets is an issue of fundamental importance in spintronics.
In this work, we show that tunneling magnetoresistance  can also emerge in junctions composed of
ferromagnets and time-reversal invariant topological superconductors without spin-rotation symmetry.
Here the physical origin  is that when the spin-polarization direction of injected electron
from the ferromagnet lying in the same plane of the spin-polarization direction of Majorana zero modes, the
electron will undergo a perfect spin-equal Andreev reflection, while injected electrons with other spin-polarization
direction will be partially Andreev reflected and partially normal reflected, which consequently
have a lower conductance, and therefore, the magnetoresistance
effect emerges.  Compared to conventional magnetic tunnel junctions, an unprecedented advantage of
the junctions studied here is that arbitrary high tunneling magnetoresistance can be obtained
even the magnetization of the ferromagnets are weak and the
insulating tunneling barriers are featureless. Our findings provide a new fascinating mechanism to obtain high
tunneling magnetoresistance.
\end{abstract}


\maketitle

\section{I. Introduction}

Superconducting phase and ferromagnetic phase are two
most familiar and classic symmetry-broken phases. The former one
breaks the $U(1)$ gauge symmetry, while the latter one
breaks the full spin-rotation symmetry (SRS) down to an axis-fixed
rotation symmetry, as a consequence of symmetry breaking, the order parameter characterizing
the superconductor (SC) takes a fixed phase, while
the one characterizing the ferromagnet (FM) takes a fixed direction.
Interestingly, the break down of the two symmetries has
direct impact on the tunneling behavior of junctions
composed of SCs or FMs.
For junctions formed by two SCs sandwiching a thin
insulator (also known as insulating tunneling barrier (ITB))
(SC-I-SC junction), the tunneling current
depends on the phase-difference~\cite{Josephson:1962}, while for
junctions formed by two FMs sandwiching a thin
insulator (FM-I-FM junction, known as magnetic tunnel junction
(MTJ)), tunneling current depends on the angle-difference
of the magnetization directions~\cite{Julliere:1975}. The former phenomenon is known
as Josephson effect, while the latter one is known as magnetic
valve effect (MVE)~\cite{Slonczewski:1989}. Both effects are very fascinating and have
very wide applications, for the latter one, there is a quantity
named as tunneling magnetoresistance (TMR) to characterize it.
Higher TMR has always been pursed since the concept was proposed
because a higher TMR implies a better performance of the effect
in real applications, such as field sensor and magnetic
random access memory~\cite{Prinz:1998, Park:1999, Zhu:2006, Chappert:2007, Ikeda:2007, Fert:2008}.

SCs of nontrivial topological properties are known as
topological superconductors (TSCs)~\cite{Qi:2011}. Due to hosting the Majorana zero modes~\cite{Kitaev:2001,Eilliott:2015}
which have potential application in topological quantum computation~\cite{Kitaev:2003,Nayak:2008},
TSC and topological superfluid (TSF) have been
among the central themes of both condensed matter and cold atom
physics in recent years~\cite{Fu:2008, Tanaka:2009, Sau:2010,Lutchyn:2010,
Oreg:2010, Alicea:2010, Qi:2010, Potter:2010, Stanescu:2011, Mourik:2012,
Das:2012, Deng:2012, Finck:2013, Rokhinson:2012, Nadj-Perge:2014, Tewari:2007, Zhang:2008, Sato:2009,
Jiang:2011, Diel:2011, Wang:2015}. According to the existence or absence
of time-reversal symmetry, particle-hole symmetry and sublattice symmetry
(or chiral symmetry), the TSCs can be classified
in a ten-fold way~\cite{Schnyder:2008,Kitaev:2009,Ryu:2010}. It is interesting to find
that TSCs in some classes, e.g. BDI class and DIII class, also break the SRS.
Based on this observation, it leads us to expect that a FM-I-TSC junction will also
exhibit TMR if the TSC breaks SRS. A direct investigation confirms our expectation
and what interesting is that the nontrivial topological property of the SC endows
nontrivial property to the TMR. For example, we find that for a generalized
FM-I-FM-I-TSC junction, arbitrary high TMR can be obtained
even the magnetization of the ferromagnets are weak and the
insulating tunneling barriers  are featureless.

The paper is organized as follows, in Sec.II, the theoretical
model and main picture are given. In Sec.III and Sec.IV,
the tunneling spectroscopies of a one-dimensional FM-I-TSC junction and a
one-dimensional FM-I-FM-I-TSC junction are studied in detail, and based on the
tunneling spectroscopies, TMR's dependence on parameters are obtained.
In Sec.V, the higher dimensional case of the FM-I-FM-I-TSC junction
is studied, and similar results like the one-dimensional case are obtained.
In Sec.VI, discussions and conclusions about the results obtained in
previous sections are given.

\begin{figure}[htbp]
\subfigure{\includegraphics[width=4cm, height=4cm]{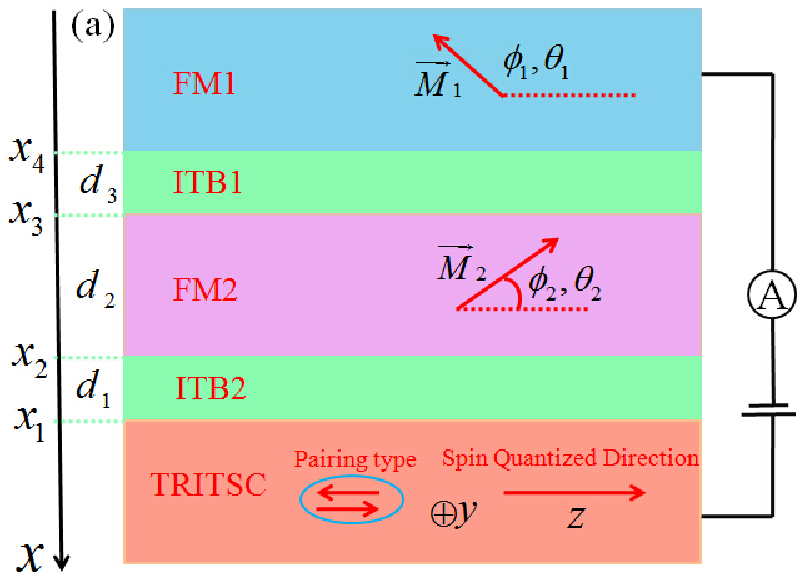}}
\subfigure{\includegraphics[width=4cm, height=4cm]{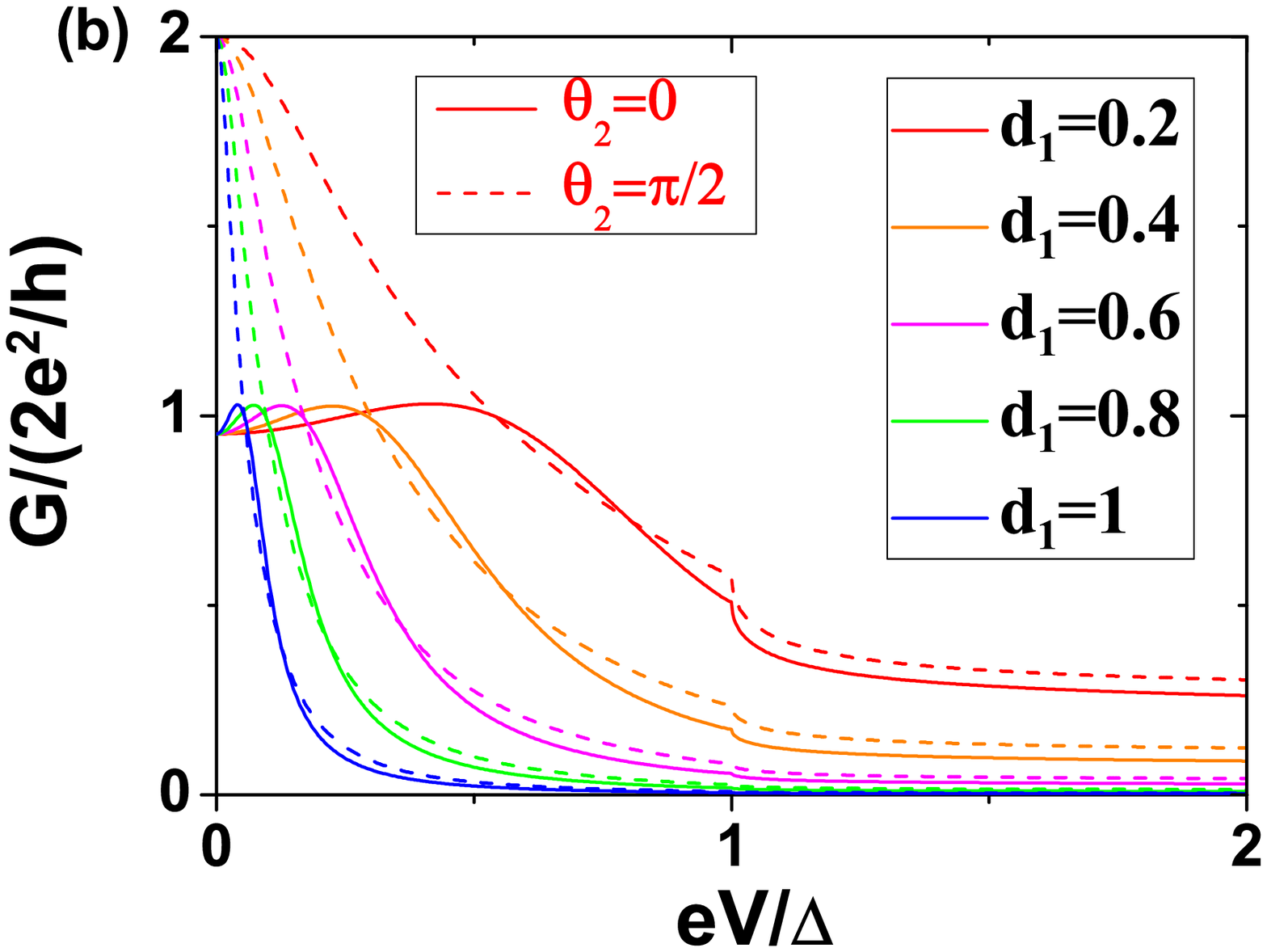}}
\subfigure{\includegraphics[width=4cm, height=4cm]{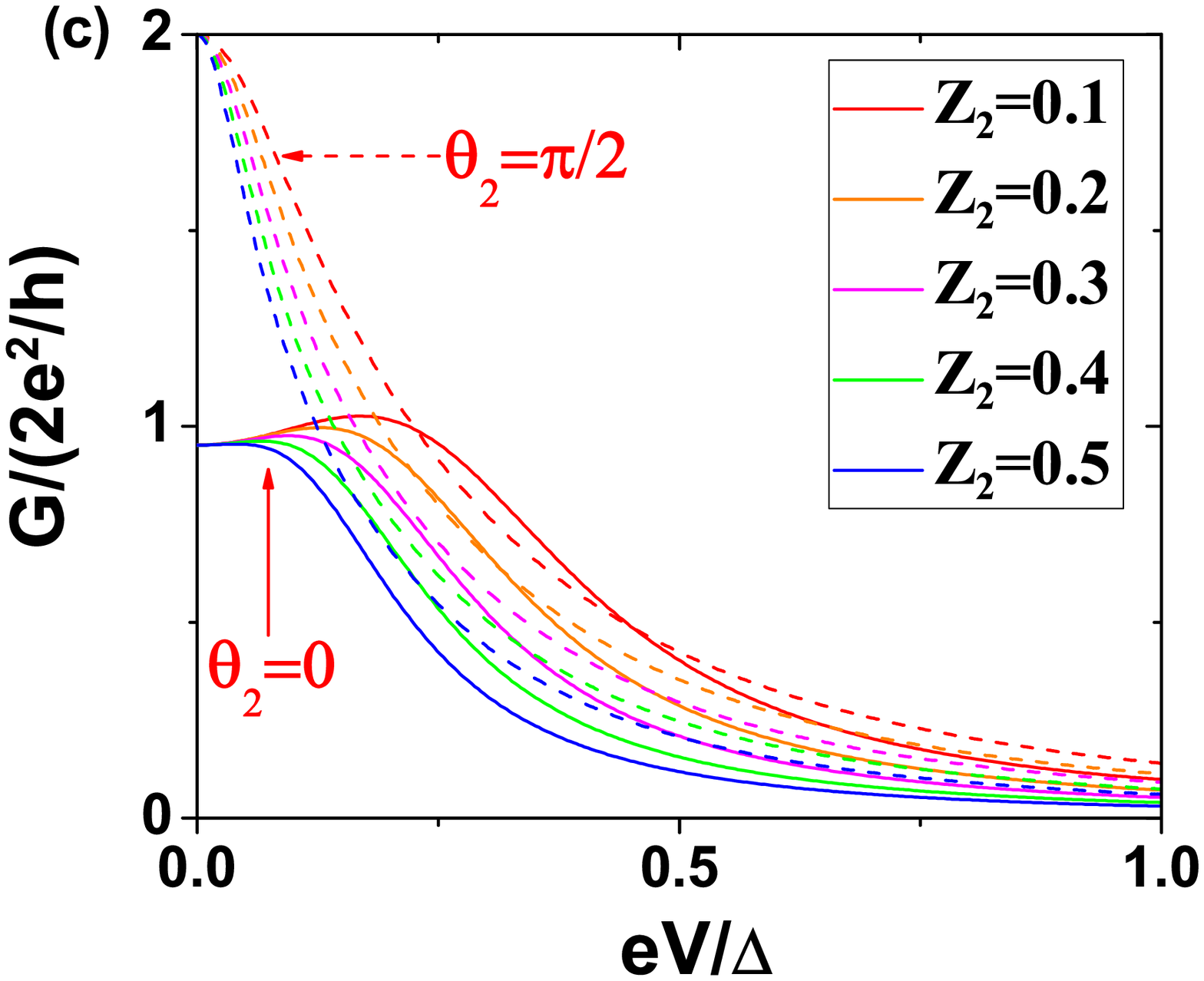}}
\subfigure{\includegraphics[width=4cm, height=4cm]{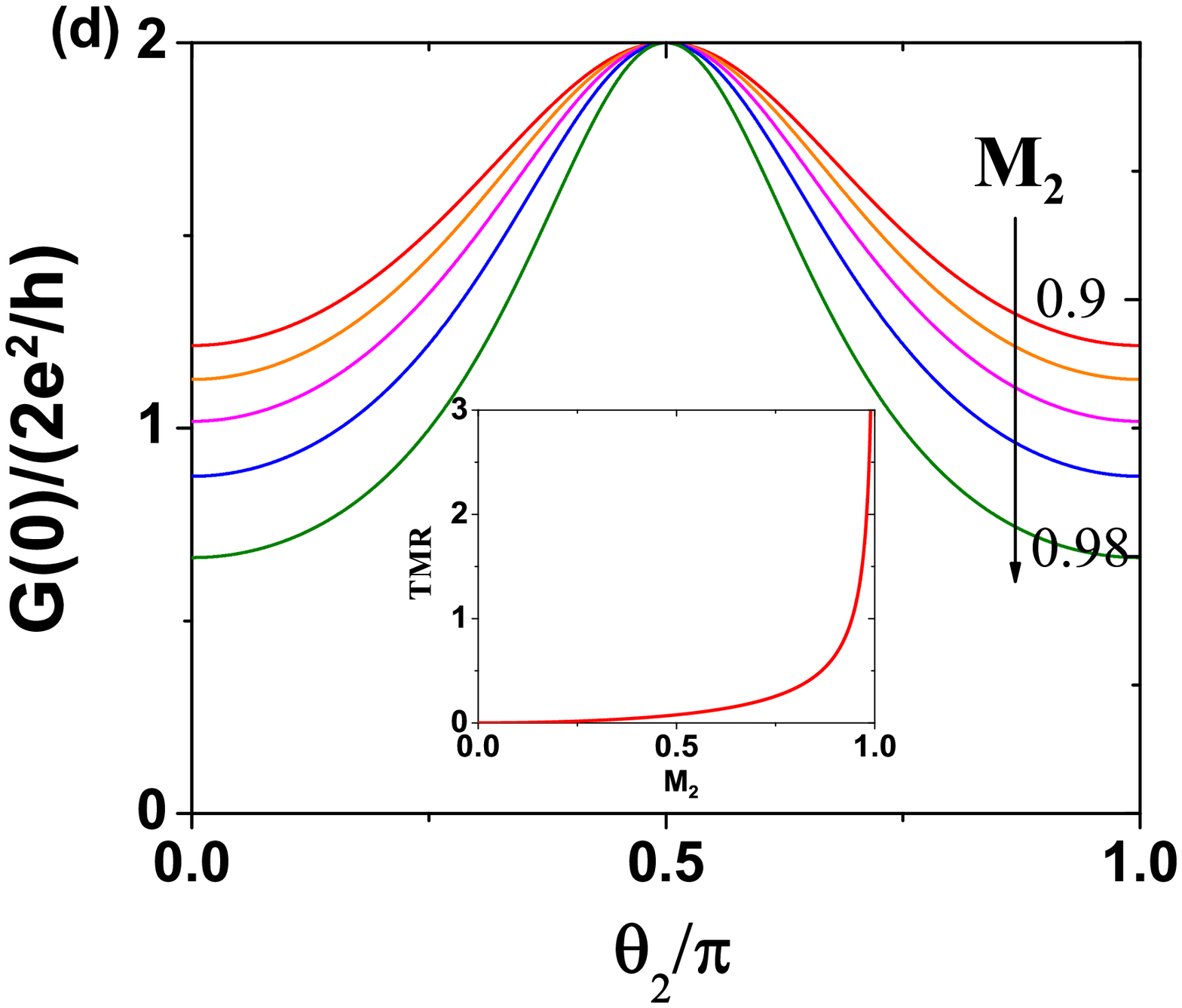}}
\caption{ Sketch of the system and tunneling spectroscopy of FM-I-TS junction. (a) The spin quantization axis of the time-reversal invariant
TSC (TRITSC) is chosen along the $z$ direction, the magnetization direction of FM1 and FM2 are characterized by $(\phi_{1},\theta_{1})$ and
$(\phi_{2},\theta_{2})$, respectively. Common parameters in (b)(c)(d), $\mu_{f}=\mu_{I}=\mu_{s}=1$ ($\mu_{f}$
is taken as the energy unit), $\Delta_{0}=0.01$, other parameters:
(b) $M_{2}=0.95$, $Z_{1}=Z_{2}=0.1$, $\Delta$ is the energy gap of the TSC.
(c) $M_{2}=0.95$, $d_{1}=0.5$, $Z_{1}=0.1$. (d) The dependence of ZBC on $\theta_{2}$ and magnetization strength, the inset shows
the TMR's dependence on  magnetization strength.
\label{fig1}}
\end{figure}

\section{II. The theoretical model and main picture}

The general picture of the junction structure we will study is
illustrated in Fig.\ref{fig1}(a). But to obtain a clear physical picture,
we will first study the simplest case, a one-dimensional FM-I-TSC junction.
Before we write down the Hamiltonian, it is worth stressing the fact
that in one dimension, if the Cooper pairs of the superconductor is spin-polarized ($S_{z}\neq0$, $\vec{S}$ is the
total spin angular momentum of the Cooper pair), the zero-bias conductance (ZBC)
of a FM-I-TSC is always quantized no matter what magnetization direction the FM chooses. Therefore, to obtain
remarkable TMR, the TSC needs to be time-reversal invariant and the Cooper pairs need to
be un-spin-polarized ($S_{z}=0$)~\cite{Wong:2012, Zhang:2013, Sato:2011, Tanaka:2012}.

Although we plan to study the one-dimensional FM-I-TSC junction firstly, here for compactness, we will
write down the general Hamiltonian describing the system illustrated in Fig.\ref{fig1}(a). Under
the representation $\hat{\Psi}^{\dag}(x)
=(\hat{\psi}^{\dag}_{\uparrow}(x),\hat{\psi}_{\downarrow}(x),\hat{\psi}^{\dag}_{\downarrow}(x),
\hat{\psi}_{\uparrow}(x))$, the general one-dimensional (higher dimensions will be studied in
lateral section) Hamiltonian is given as ($\hbar=m=1$) \cite{Yan:2015}
\begin{eqnarray}
\mathcal{H}=\tau_{z}\left[-\frac{\partial^{2}_{x}}{2} - \mu(x) + V(x)
\right] + \tau_{x}\Delta(x), \label{1}
\end{eqnarray}
where $\vec{\tau}=(\tau_{x}, \tau_{y}, \tau_{z})$ are Pauli matrices
in particle-hole space, $V(x)$ is potential induced by disorder, external
field, $etc$, here we assume it takes the form
\begin{eqnarray}
&&V(x)=-\tau_{z}M_{1}(\vec{n}_{1}\cdot\vec{\sigma})\Theta(x_{3}-x)-\tau_{z}M_{2}(\vec{n_{2}}\cdot\vec{\sigma})\nonumber\\
&&\qquad\Theta(x-x_{2})\Theta(x_{1}-x)+\sum_{i=1}^{4}Z_{i}\delta(x_{i}-x),\label{2}
\end{eqnarray}
the terms in the first line denote the magnetization of FM1 and FM2, $M_{1}$
and $M_{2}$ denote the magnetization strength of FM1 and FM2, while $\vec{n}_{1}=(\cos\phi_{1}\sin\theta_{1},
\sin\phi_{1}\sin\theta_{1},\cos\theta_{1})$ and $\vec{n}_{2}=(\cos\phi_{2}\sin\theta_{2},
\sin\phi_{2}\sin\theta_{2},\cos\theta_{2})$ denote the magnetization directions
of FM1 and FM2, respectively; $\sigma_{x,y,z}$
are Pauli matrices acting on the spin space, $\Theta(x)$ is the Heaviside function,
$i.e.$, $\Theta(x>0)=1$, $\Theta(x<0)=0$;
the terms in the second line denote the scattering potential at the interfaces; $\mu(x)$ is
the chemical potential, we set $\mu(x)=\mu_{f}$ (or $E_{F}$) for the two FMs,
$\mu(x)=-\mu_{I}$ for the two insulating tunneling regions, and $\mu(x) = \mu_{s}>0$ for the SC.
$\Delta(x)=-i\Delta_{0}\Theta(x-x_{1})\partial_{x}$ is the pairing potential, which is assumed to be
$p$-wave type (then as $\mu_{s}>0$ is assumed, the SC is a TSC with Majorana zero modes
located at the boundary~\cite{Wan:2014, Bernevig:2013}) and homogeneous at $x>x_{1}$ and vanish
at $x<x_{1}$ for the sake of theoretical simplicity.

To see that the TSC breaks the SRS, we transform the Hamiltonian corresponding to the
superconducting part into momentum space, then under the representation $\hat{\Psi}^{\dag}(k)
=(\hat{\psi}^{\dag}_{\uparrow}(k),\hat{\psi}_{\downarrow}(-k),\hat{\psi}^{\dag}_{\downarrow}(k),
\hat{\psi}_{\uparrow}(-k))$
\begin{eqnarray}
\mathcal{H}_{sc}(k)=[\frac{k^{2}}{2}
-\mu_{s}]\tau_{z}\sigma_{0}+\Delta(k)\tau_{x}\sigma_{0}. \label{3}
\end{eqnarray}
where $\Delta(k)=\Delta_{0}k$, $\sigma_{0}$ is the $2\times2$ unit matrix in spin space.
By making a spin-rotation:
$\hat{\psi}_{\uparrow}(k)\rightarrow\cos\theta\hat{\psi}_{\uparrow}(k)+\sin\theta\hat{\psi}_{\downarrow}(k)$,
$\hat{\psi}_{\downarrow}(k)\rightarrow-\sin\theta\hat{\psi}_{\uparrow}(k)+\cos\theta\hat{\psi}_{\downarrow}(k)$,
it is easy to check that the superconducting term, $\Delta_{0}k\hat{\psi}^{\dag}_{\uparrow}(k)\hat{\psi}^{\dag}_{\downarrow}(-k)
+h.c.$, is not invariant, and therefore breaks the SRS. The time-reversal symmetry of the Hamiltonian is easy to check: $\mathcal{T}\mathcal{H}_{sc}(k)\mathcal{T}^{-1}=\mathcal{H}_{sc}(-k)$ with $\mathcal{T}=\tau_{z}K$, where
$K$ is the complex conjugate operator.

To obtain the tunneling spectroscopies of the junction, here we follow the
Slonczewski~\cite{Slonczewski:1989} and Blonder-Tinkham-Klapwijk (BTK)
approach~\cite{Blonder:1982, Tanaka:1995, Tanaka:2000}. The first step of the approach is to write down
the wave functions of each part of the junctions. If we consider that a
majority electron with energy $E$ (relative to $\mu_{f}$) is injected from FM1,
the wave function in FM1 is given as
\begin{eqnarray}
&&\psi_{FM1}=\vec{\chi}_{1}e^{ik_{1+,e}x}+b_{1+}\vec{\chi}_{1}e^{-ik_{1+,e}x}+
a_{1+}\vec{\chi}_{2}e^{ik_{1+,h}x}\nonumber\\
&&\qquad\quad+b_{1-}\vec{\chi}_{3}e^{-ik_{1-,e}x}
+a_{1-}\vec{\chi}_{4}e^{ik_{1-,h}x} \label{4}
\end{eqnarray}
with
\begin{eqnarray}
&&\vec{\chi}_{1}=(\eta_{1},0,\eta_{2}e^{i\phi_{1}},0)^{T},
\vec{\chi}_{2}=(0,\eta_{2},0,\eta_{1}e^{i\phi_{1}})^{T}, \nonumber\\
&&\vec{\chi}_{3}=(-\eta_{2},0,\eta_{1}e^{i\phi_{1}},0)^{T},
\vec{\chi}_{4}=(0,-\eta_{1},0,\eta_{2}e^{i\phi_{1}})^{T},\label{5}
\end{eqnarray}
where $\eta_{1}=\cos(\theta_{1}/2)$, $\eta_{2}=\sin(\theta_{1}/2)$ and
$k_{1+,e(h)}=\sqrt{2(\mu_{f}+M_{1}+(-)E)}$, $k_{1-,e(h)}=\sqrt{2(\mu_{f}-M_{1}+(-)E)}$.
The coefficients $b_{1+(-)}$ and $a_{1+(-)}$ denote the
spin-equal (spin-opposite) normal reflection (a majority electron reflected as a majority
(minority) electron) amplitude and spin-equal (spin-opposite) Andreev
reflection (a majority electron reflected as a majority (minority) hole~\cite{Andreev:1964})
amplitude, respectively. The wave functions in other parts can be obtained easily but as
their forms are tedious, their concrete expressions will be given in the Appendix.

To obtain the tunneling conductance, the coefficients in the wave functions need to
be determined by matching the wave functions at the interfaces according to
the boundary conditions~\cite{Slonczewski:1989,Blonder:1982, Tanaka:1995, Tanaka:2000}
\begin{eqnarray}
&&\psi_{L}(x=x_{i})=\psi_{R}(x=x_{i}),\nonumber\\
&&v_{i,R}\psi_{R}(x_{i}^{+})-v_{i,L}\psi_{L}(x_{i}^{-})=-i2Z_{i}\tau_{z}\sigma_{0}\psi_{R}(x_{i}),\label{6}
\end{eqnarray}
where $\psi_{L}(x=x_{i})$ ($\psi_{R}(x=x_{i})$) denotes the wave function in the left (right) neighbouring
part of the interface located at $x=x_{i}$, $v_{i,R(L)}=\partial \mathcal{H}_{i,R(L)}/\partial k$ is the
velocity operator corresponding to the right (left) neighboring part of the interface~\cite{Wan:2014}.

After the coefficients are obtained, the zero temperature tunneling conductance can be determined according
to the BTK formula~\cite{Blonder:1982}
\begin{eqnarray}
G_{+}(E=eV,\vec{n}_{1},\vec{n}_{2})=\frac{e^{2}}{h}(1+A_{+}+A_{-}
-B_{+}-B_{-}),\label{7}
\end{eqnarray}
where $A_{+}=k_{1+,h}|a_{n+}|^{2}/k_{1+,e}$,
$B_{+}=|b_{1+}|^{2}$,
$A_{-}=k_{1-,h}|a_{1-}|^{2}/k_{1+,e}$,
$B_{-}=k_{1-,e}|b_{1-}|^{2}/k_{1+,e}$.
Similar procedures can obtain $G_{-}(eV,\vec{n}_{1},\vec{n}_{2})$, the tunneling
conductance for a minority electron, and the total
tunneling conductance $G(eV,\vec{n}_{1},\vec{n}_{2})$ is given as the summation of
$G_{+}(eV,\vec{n}_{1},\vec{n}_{2})$ and $G_{-}(eV,\vec{n}_{1},\vec{n}_{2})$.

\section{III. One-dimensional FM-I-TSC  junction}
\label{III}

Now, we are going to study the one-dimensional FM-I-TSC junction.
For the FM-I-TSC  junction, we consider that it is only composed
of FM2, ITB2 and TSC, and $x_{3}$ is set to infinity.
For this structure, the wave function in FM2 takes the same form as $\psi_{FM1}$,
but with a substitution of the parameters:
$(k_{1+(-),e(h)},\theta_{1},\phi_{1})\rightarrow(k_{2+(-),e(h)},\theta_{2},\phi_{2})$,
where $k_{2+,e(h)}=\sqrt{2(\mu_{f}+M_{2}+(-)E)}$, $k_{2-,e(h)}=\sqrt{2(\mu_{f}-M_{2}+(-)E)}$.

\begin{figure}[t]
\subfigure{\includegraphics[width=4cm, height=3.5cm]{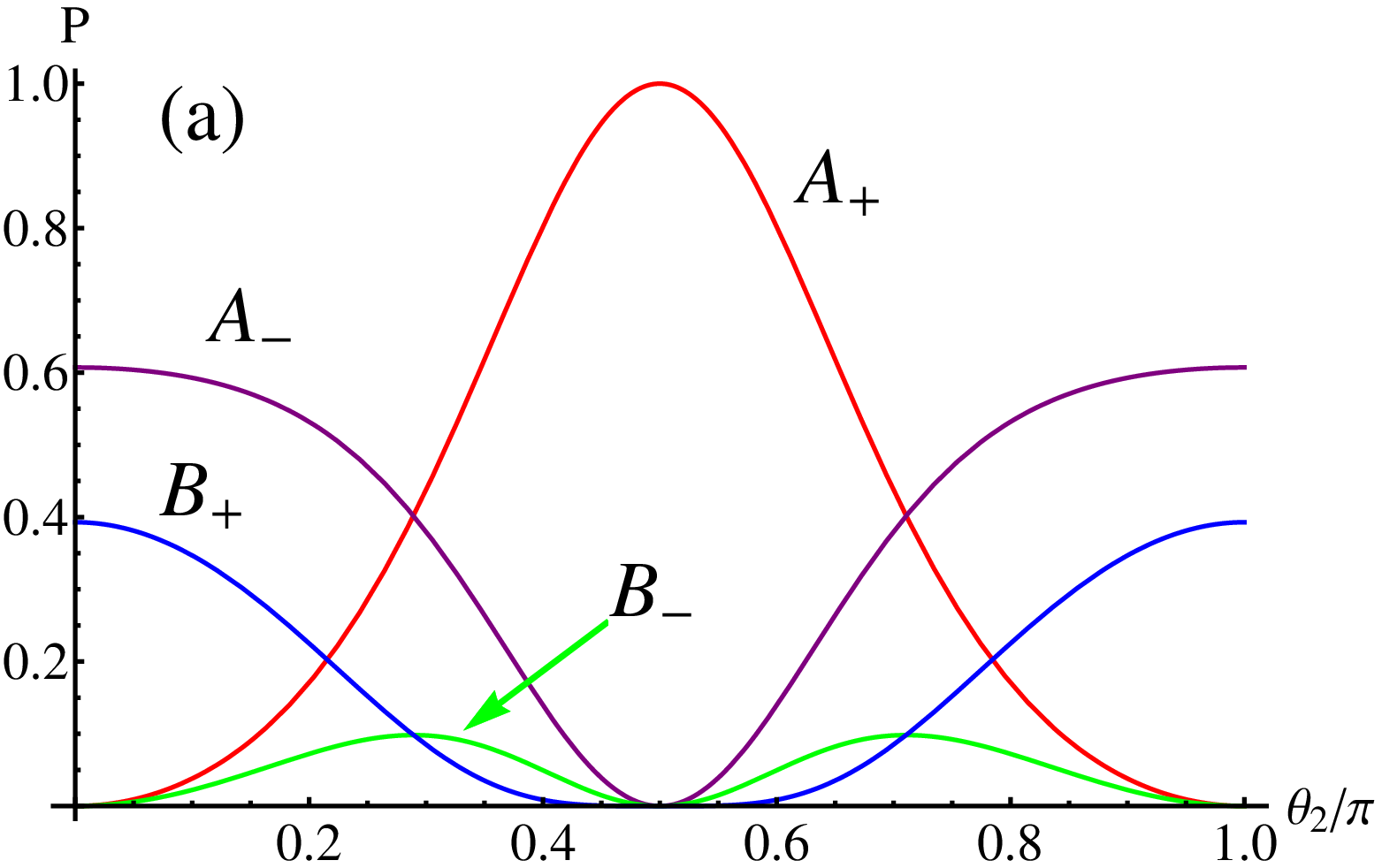}}
\subfigure{\includegraphics[width=4cm, height=3.5cm]{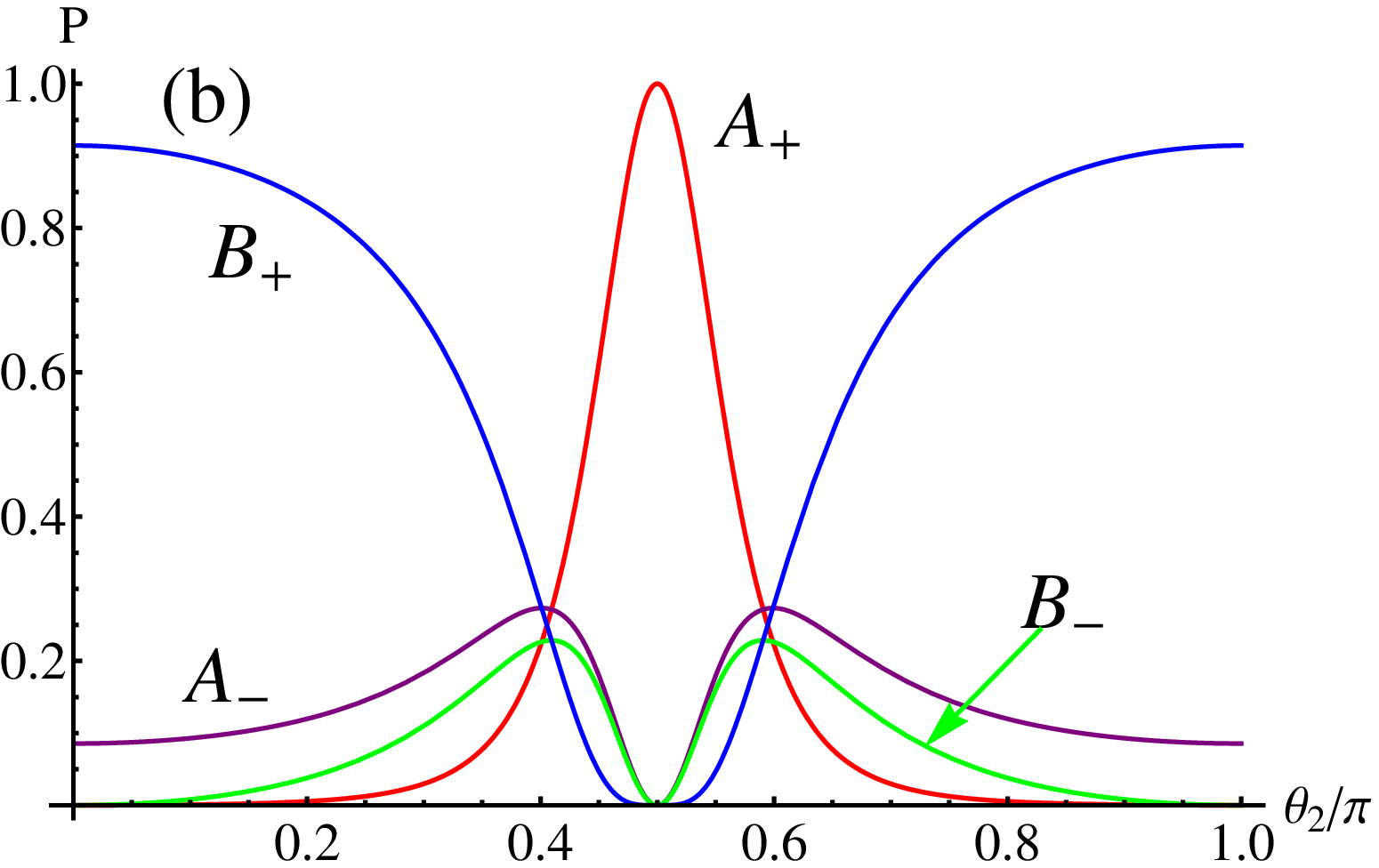}}
\subfigure{\includegraphics[width=4cm, height=4cm]{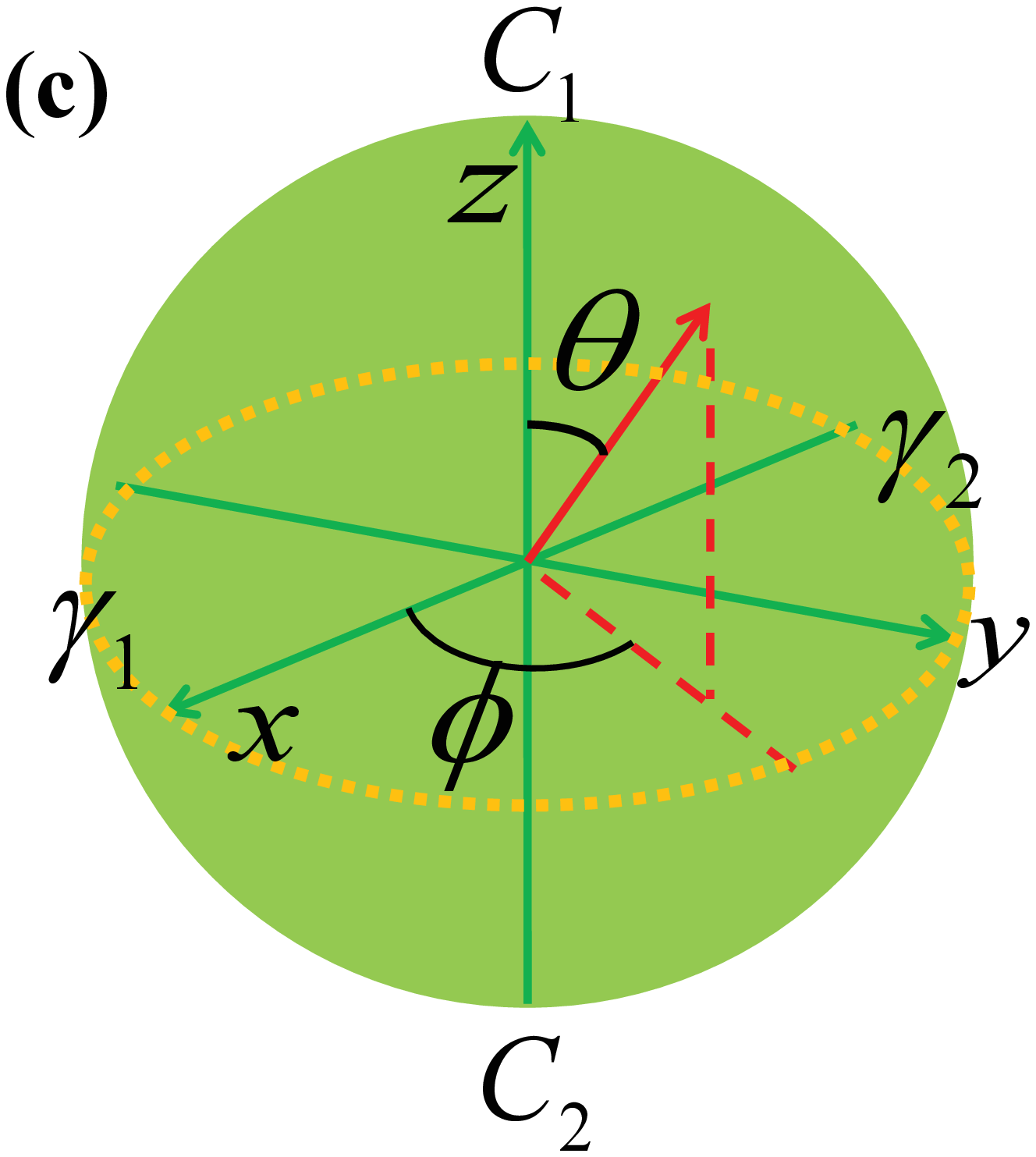}}
\subfigure{\includegraphics[width=4cm, height=4cm]{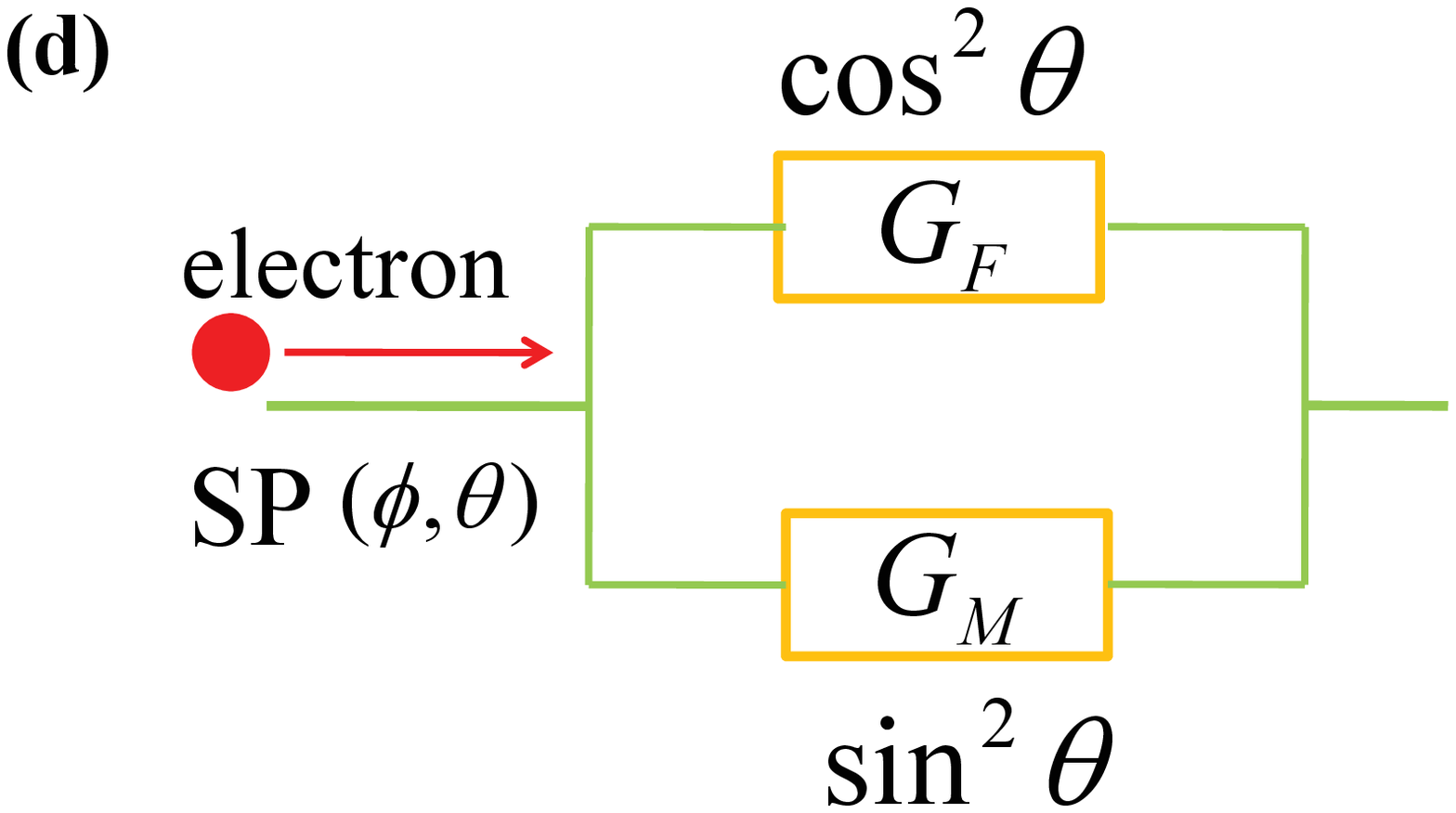}}
\caption{ The four reflection coefficients with (a) $M_{2}/\mu_{f}=0.9$ and (b) $M_{2}/\mu_{f}=0.9999$.
(c) Spin-polarization direction of the zero modes. (d) A circuit picture to understand
the tunneling process and the result given in eq.(\ref{9}).}
\label{fig2}
\end{figure}

By a simple numeric calculation of the coefficients of the wave functions, the tunneling
conductance of the FM-I-TSC junction is shown in Fig.\ref{1}(b)(c).
There are two extraordinary characteristics in the tunneling spectroscopy. The first one is that the tunneling
conductance is angle-dependent, the second one is that the tunneling conductance at zero-bias voltage is
of topological feature in the sense that it is independent of the thickness of ITB2 and the interface scattering
potential. In fact, at zero-bias voltage,  the four key quantities $A_{+}$, $A_{-}$,
$B_{+}$ and $A_{-}$ can be analytically obtained. When we consider that a majority
 electron is injected, their analytical forms are
\begin{eqnarray}
&&A_{+}=\frac{16k_{2+}^{2}k_{2-}^{2}\sin^{2}\theta_{2}}{[(k_{2+}+k_{2-})^{2}\cos^{2}\theta_{2}
+4k_{2+}k_{2-}\sin^{2}\theta_{2}]^{2}},\nonumber\\
&&A_{-}=\frac{4k_{2+}k_{2-}(k_{2+}+k_{2-})^{2}\cos^{2}\theta_{2}}{[(k_{2+}+k_{2-})^{2}\cos^{2}\theta_{2}
+4k_{2+}k_{2-}\sin^{2}\theta_{2}]^{2}},\nonumber\\
&&B_{+}=\frac{(k_{2+}^{2}-k_{2-}^{2})^{2}\cos^{4}\theta_{2}}{[(k_{2+}+k_{2-})^{2}\cos^{2}\theta_{2}
+4k_{2+}k_{2-}\sin^{2}\theta_{2}]^{2}},\nonumber\\
&&B_{-}=\frac{4k_{2+}k_{2-}(k_{2+}-k_{2-})^{2}\sin^{2}
\theta_{2}\cos^{2}\theta_{2}}{[(k_{2+}+k_{2-})^{2}\cos^{2}\theta_{2}
+4k_{2+}k_{2-}\sin^{2}\theta_{2}]^{2}},\label{8}
\end{eqnarray}
where $k_{2+(-)}=\sqrt{2(\mu_{f}+(-)M_{2})}$. If a minority electron is injected,
the four key quantities have such an exchange, $A_{+}\leftrightarrow A_{-}$,
$B_{+}\leftrightarrow B_{-}$. Then according to the eq.(\ref{7}),
it is direct to obtain
\begin{eqnarray}
G(0,\vec{n}_{2})=\frac{e^{2}}{h}\frac{16k_{2+}k_{2-}}
{(k_{2+}+k_{2-})^{2}\cos^{2}\theta_{2}+4k_{2+}k_{2-}\sin^{2}\theta_{2}}.\label{9}
\end{eqnarray}
The zero-bias conductance (ZBC) only depends on $\theta_{2}$ and the parameters of FM2.
From the formula we can see that $G(0,\vec{n}_{2})$ is quantized as $4e^{2}/h$ at
$\theta_{2}=\pi/2$ and  takes
its minimum value at $\theta_{2}=0$ and $\theta_{2}=\pi$, as shown in fig.\ref{1}(d).
In a recent work~\cite{Yan:2015}, we have proven  that by making use of this
minimum value, it is very convenient to determine the polarization of FMs.
From eq.(\ref{8})
or more directly from fig.\ref{fig2}(a)(b),
we can see that at $\theta_{2}=\pi/2$, $A_{+}=1$, while the other three quantities
are all equal to zero, which indicates a perfect spin-equal Andreev reflection.
As this perfect spin-equal Andreev reflection is unaffected by the ITB2 and the scattering
potential, here the only possible origin is the resonant tunneling due to the topological
Majorana zero modes located at the boundary of the TSC. In ref.\cite{He:2014},
by using scattering matrix, the authors have shown that perfect spin-equal Andreev reflection will
occur when the injected electron takes certain spin-polarization direction. In the following,
we will show that here the magic direction is just the
spin-polarization direction of the Majorana zero modes.

To obtain the zero modes in the TSC, we need to calculate the BdG equation, which
is given as
\begin{eqnarray}
&&Eu_{\uparrow}(x)=(-\frac{\partial_{x}^{2}}{2}-\mu)u_{\uparrow}(x)-i\Delta\partial_{x}v_{\downarrow}(x),\nonumber\\
&&Ev_{\downarrow}(x)=-i\Delta\partial_{x}u_{\uparrow}(x)+(\frac{\partial_{x}^{2}}{2}+\mu)v_{\downarrow}(x),\nonumber\\
&&Eu_{\downarrow}(x)=(-\frac{\partial_{x}^{2}}{2}-\mu)u_{\downarrow}(x)-i\Delta\partial_{x}v_{\uparrow}(x),\nonumber\\
&&Ev_{\uparrow}(x)=-i\Delta\partial_{x}u_{\downarrow}(x)+(\frac{\partial_{x}^{2}}{2}+\mu)v_{\uparrow}(x),\label{10}
\end{eqnarray}
with the boundary condition that the wave functions $u_{\uparrow}(x)$,
$v_{\downarrow}(x)$, $u_{\downarrow}(x)$ and $v_{\uparrow}(x)$ all need
to vanish at $x=0$ and $x=+\infty$.
For the zero modes, $i.e.$, $E=0$, a direct calculation gives
\begin{eqnarray}
&&u_{\uparrow}(x)=iv_{\downarrow}(x)=\mathcal{N}\sin(\kappa x)e^{-\gamma x},\nonumber\\
&&u_{\downarrow}(x)=iv_{\uparrow}(x)=\mathcal{N}\sin(\kappa x)e^{-\gamma x},\label{11}
\end{eqnarray}
where $\kappa=\sqrt{2\mu-\Delta^{2}}$, $\gamma=\Delta$, and $\mathcal{N}$ is
the normalization constant which guarantees $\int_{0}^{+\infty} dx(|u_{,\uparrow(\downarrow)}(x)|^{2}+|v_{\downarrow(\uparrow)}(x)|^{2})=1$,
then the zero modes located at the $x=0$ boundary is given as
\begin{eqnarray}
&&C_{1}=|\mathcal{N}|\int dx [e^{\frac{i\pi}{4}}\psi_{\uparrow}(x)+e^{-\frac{i\pi}{4}}\psi_{\downarrow}^{\dag}(x)]\sin(\kappa x)e^{-\gamma x},\nonumber\\
&&C_{2}=|\mathcal{N}|\int dx [e^{\frac{i\pi}{4}}\psi_{\downarrow}(x)+e^{-\frac{i\pi}{4}}\psi_{\uparrow}^{\dag}(x)]\sin(\kappa x)e^{-\gamma x},\label{12}
\end{eqnarray}
here we have chosen $\mathcal{N}=|\mathcal{N}|e^{i\pi/4}$ for the convenience
of the following discussions. By using the normalization condition, it is direct to
verify $\{C_{1}, C_{1}^{\dag}\}=\{C_{2}, C_{2}^{\dag}\}=1$. It is also easy to see that
$C_{1}$ and $C_{2}$ are related with each other
by the charge conjugate, $i.e.$, $C_{1}^{\dag}=C_{2}$, but $C_{1}^{\dag}\neq C_{1}$
and $C_{2}^{\dag}\neq C_{2}$, this indicates that the two fermionic zero modes $C_{1}$ and $C_{2}$ are
not of Majorana characteristic. However, from the Kitaev Majorana chain
model we already know that a fermion operator with its conjugate can construct
two Majorana operators~\cite{Kitaev:2001}, here as $C_{1}^{\dag}=C_{2}$, the two Majorana zero
modes can be constructed as
\begin{eqnarray}
\gamma_{1}=\frac{C_{1}+C_{1}^{\dag}}{\sqrt{2}}, \gamma_{2}=\frac{C_{1}-C_{1}^{\dag}}{\sqrt{2}i},\label{13}
\end{eqnarray}
the Majorana zero modes are an equal-weight superposition of the two fermionic zero modes.

Based on eq.(\ref{12}) and eq.(\ref{13}), the spin-polarization of the zero modes can be directly calculated,
\begin{eqnarray}
<C_{1}|\hat{S}_{x}|C_{1}> &=& <C_{1}|\hat{S}_{y}|C_{1}>=0, <C_{1}|\hat{S}_{z}|C_{1}>=\frac{\hbar}{2},\nonumber\\
<C_{2}|\hat{S}_{x}|C_{2}> &=& <C_{2}|\hat{S}_{y}|C_{2}>=0, <C_{2}|\hat{S}_{z}|C_{2}>=-\frac{\hbar}{2},\nonumber\\
<\gamma_{1}|\hat{S}_{y}|\gamma_{1}> &=& <\gamma_{1}|\hat{S}_{z}|\gamma_{1}>=0, <\gamma_{1}|\hat{S}_{x}|\gamma_{1}>=\frac{\hbar}{2},\nonumber\\
<\gamma_{2}|\hat{S}_{y}|\gamma_{2}> &=& <\gamma_{2}|\hat{S}_{z}|\gamma_{2}>=0, <\gamma_{2}|\hat{S}_{x}|\gamma_{2}>=-\frac{\hbar}{2},\qquad \label{14}
\end{eqnarray}
where $|C_{1}>=(e^{-\frac{i\pi}{4}}(i\psi_{\uparrow}+\psi_{\downarrow}^{\dag}))^{\dag}|0>$,
$|C_{2}>=(e^{-\frac{i\pi}{4}}(i\psi_{\downarrow}+\psi_{\uparrow}^{\dag}))^{\dag}|0>$,
$|\gamma_{1}>=(e^{-\frac{i\pi}{4}}/\sqrt{2})(i\psi_{\uparrow}+\psi_{\downarrow}^{\dag}+i\psi_{\downarrow}+\psi_{\uparrow}^{\dag})|0>$,
$|\gamma_{1}>=(e^{-\frac{i\pi}{4}}/\sqrt{2}i)(i\psi_{\uparrow}+\psi_{\downarrow}^{\dag}-i\psi_{\downarrow}-\psi_{\uparrow}^{\dag})|0>$,
and $\hat{S}_{x}=\frac{\hbar}{2}(\psi_{\uparrow}^{\dag}\psi_{\downarrow}+\psi_{\downarrow}^{\dag}\psi_{\uparrow})$,
$\hat{S}_{y}=\frac{-i\hbar}{2}(\psi_{\uparrow}^{\dag}\psi_{\downarrow}-\psi_{\downarrow}^{\dag}\psi_{\uparrow})$,
$\hat{S}_{z}=\frac{\hbar}{2}(\psi_{\uparrow}^{\dag}\psi_{\uparrow}-\psi_{\downarrow}^{\dag}\psi_{\downarrow})$.
From eq.(\ref{14}),  it is direct to see that
the spin-polarization directions of the two fermionic zero modes $C_{1}$ and $C_{2}$ are along
the positive $z$-direction and negative $z$-direction, respectively, while the spin-polarization of the two
Majorana zero modes $\gamma_{1}$ and $\gamma_{2}$ are along the positive $x$-direction and negative
$x$-direction, respectively, which is a natural result according to eq.(\ref{13}). Projecting the
spin-polarization on the Bloch sphere, the spin-polarization directions of the two fermionic
zero modes $C_{1}$ and $C_{2}$ point to the north pole ($\theta_{2}=0$) and south pole ($\theta_{2}=\pi$), respectively, while
the spin-polarization directions of the two Majorana zero modes are lying in the equatorial plane,
which corresponds to $\theta_{2}=\pi/2$, as shown in fig.\ref{fig2}(c).

Based on these results, the physical picture is clear: when the
spin-polarization direction of the injected electron is along the
$z$-direction, $\theta_{2}=0$ or $\theta_{2}=\pi$, the injected electron only feels
one of the fermionic zero modes and the only possible Andreev reflection
is the spin-opposite Andreev reflection. For the spin-opposite
Andreev reflection, the Fermi-surface mismatch in the FM due to the magnetization will suppress it
and therefore, it can not be perfect and spin-equal normal reflection will take place, as shown
in fig.\ref{fig2}(a)(b). But as the fermionic zero modes
are of topological nature, the insulating tunneling barrier
and the interface scattering potential can not affect the spin-opposite Andreev
reflection.
When the spin-polarization direction of the injected electron
is lying in the equatorial plane, $i.e.$, $\theta_{2}=\pi/2$,
it feels the two fermionic zero modes equally and is equivalent
to couple with the Majorana zero modes, and then the only possible Andreev reflection is the
spin-equal Andreev reflection (because the electron coupled with either one of
the two Majorana zero modes will undergo a perfect spin-equal Andreev reflection, the azimuthal angle
$\phi_{2}$ has no effect to the tunneling process). For the spin-equal
Andreev reflection, however, the magnetization has no effect to it and therefore, the electron will undergo
a perfect Andreev reflection and consequently results in a quantized conductance.
For the general case that the spin-polarization direction of the injected electrons is neither along
the fermionic zero modes' nor the majorana zero modes', the electron can be divided into two parts,
one along the spin-polarization direction of the fermionic zero modes and the other
one along the Majorana zero modes'. Based on this division, the conductance given in eq.(\ref{9})
can also correspondingly be divided into two parts,
\begin{eqnarray}
\frac{1}{G(0,\vec{n}_{2})}=\frac{1}{G_{F}}\cos^{2}\theta_{2}+\frac{1}{G_{M}}\sin^{2}\theta_{2},\label{15}
\end{eqnarray}
where $G_{F}=16k_{2+}k_{2-}G_{0}/(k_{2+}+k_{2-})^{2}$ and
$G_{M}=4G_{0}$ with $G_{0}=e^{2}/h$, $G_{F}$ corresponds to the process that the injected
electron is coupled with only one of the fermionic zero modes, and $G_{M}$ corresponds to the
process that the injected electron is coupled with the Majorana zero modes.
The physical meaning of eq.(\ref{15}) is that the two processes form a parallel circuit
as shown in fig.\ref{fig2}(d). An even more clear picture can be obtained if the conductance is
substituted by resistance, $i.e.$, $R=1/G$,
then eq.(\ref{15}) is correspondingly rewritten as
\begin{eqnarray}
R(0,\vec{n}_{2})=R_{F}\cos^{2}\theta_{2}+R_{M}\sin^{2}\theta_{2}. \label{16}
\end{eqnarray}
The physical meaning of eq.(\ref{16}) is that the total
resistance of the tunneling process is a sum of the
resistances of the two possible tunneling processes,
which is obviously physical right.

As $G(0,\vec{n}_{2})$ has an explicit angle dependence,
the junction will naturally exhibit TMR.
Because low-bias voltage corresponds to low-energy consumption which is
very important for real applications, the low-bias regime is of central
interest. Therefore, in the following, when we
consider the TMR, we will restrict
ourselves to the zero-bias voltage. In the low-bias regime,
$eV<<\Delta$, where $\Delta$ is the energy gap of TSC,
the effect of increasing
the bias voltage is a reduction of the TMR, but
the reduction will be quite limited and the main physics
obtained at zero-bias voltage will still hold.

Generally,  TMR  is  defined as
\begin{eqnarray}
\mathrm{TMR}=\frac{R_\mathrm{ap}-R_\mathrm{p}}{R_\mathrm{p}}=\frac{R_\mathrm{max}-R_\mathrm{min}}{R_\mathrm{min}},\label{17}
\end{eqnarray}
where $R_\mathrm{ap}$ is the electrical resistance in the anti-parallel state,
whereas $R_\mathrm{p}$ is the resistance in the parallel state. However, here
a better definition is given as
\begin{eqnarray}
\mathrm{TMR}=\frac{G_\mathrm{max}-G_\mathrm{min}}{G_\mathrm{min}},\label{18}
\end{eqnarray}

According to the formula (\ref{18}) and (\ref{9}), the TMR of the junction is given as
\begin{eqnarray}
\mathrm{TMR}=\frac{(k_{2+}-k_{2-})^{2}}{4k_{2+}k_{2-}}.\label{19}
\end{eqnarray}
Since the ZBC is of topological feature, here the TMR is also of topological feature,
which is fundamentally different from the usual cases~\cite{Slonczewski:1989}.

In eq.(\ref{19}), with the increase of magnetization strength, $k_{2-}$ will
continue to decrease to zero, and then TMR will go to diverge, as shown
in the inset of fig.\ref{fig1}(d). The divergent behavior  indicates that if the FM
turns to be a half metal, the MVE is perfect even there is only one
FM. However,  as shown in fig.\ref{fig1}(d), the increase of TMR
is quite slow, even when $M_{2}/\mu_{f}$ reaches $90\%$, the TMR
is still smaller than $100\%$, this implies that only when the FM
is close to perfect polarization, a very high TMR can be obtained.
The request of strong magnetization will greatly limit the applicable
ferromagnetic materials and consequently reduces the novelty of the
new MVE. In the following section, we will show
that a generalized FM-I-FM-I-TSC junction overcomes this shortcoming.

\section{IV. One-dimensional FM-I-FM-I-TSC junction}
\label{IV}

\begin{figure}[t]
\subfigure{\includegraphics[width=4cm, height=4cm]{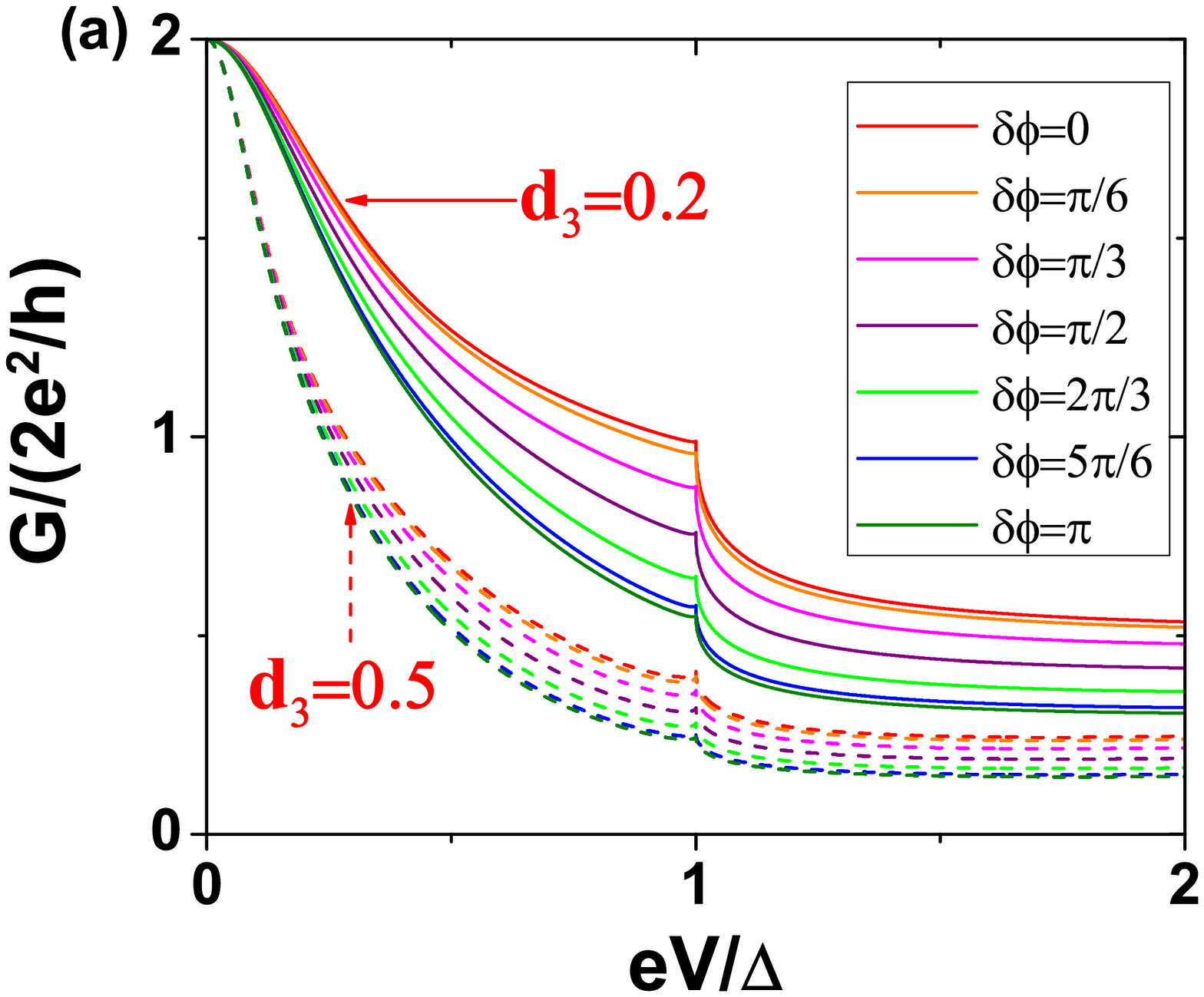}}
\subfigure{\includegraphics[width=4cm, height=4cm]{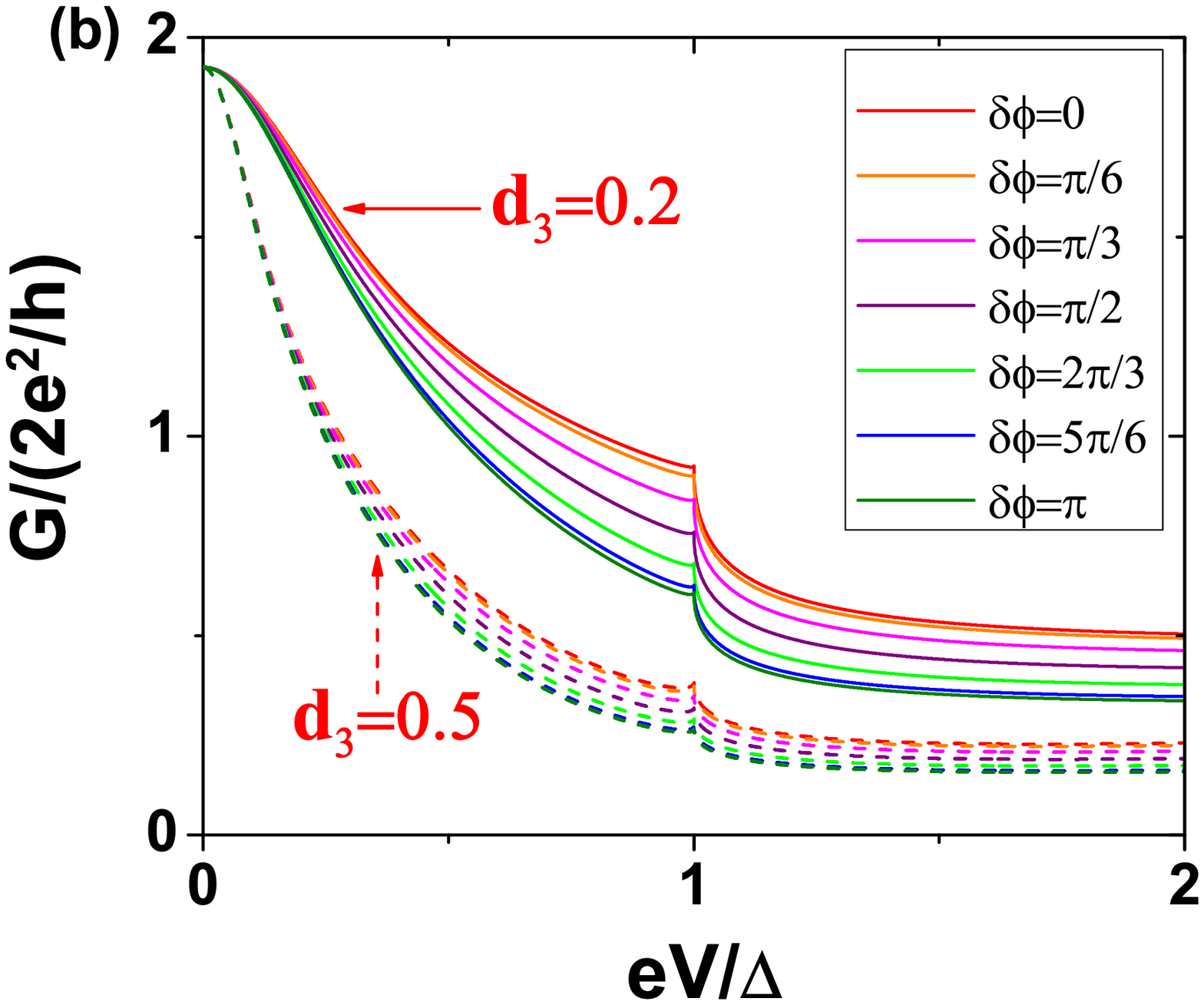}}
\subfigure{\includegraphics[width=4cm, height=4cm]{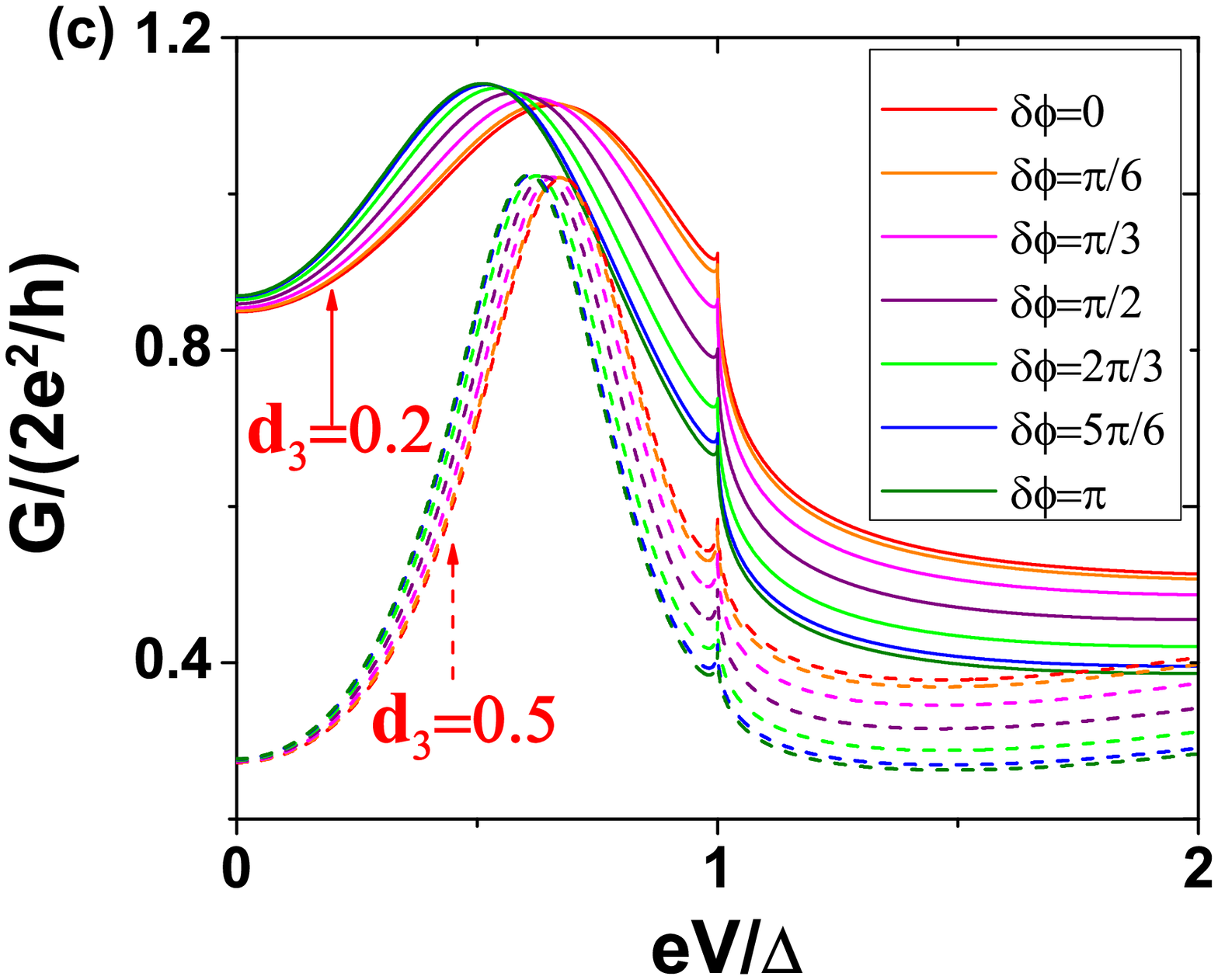}}
\subfigure{\includegraphics[width=4cm, height=4cm]{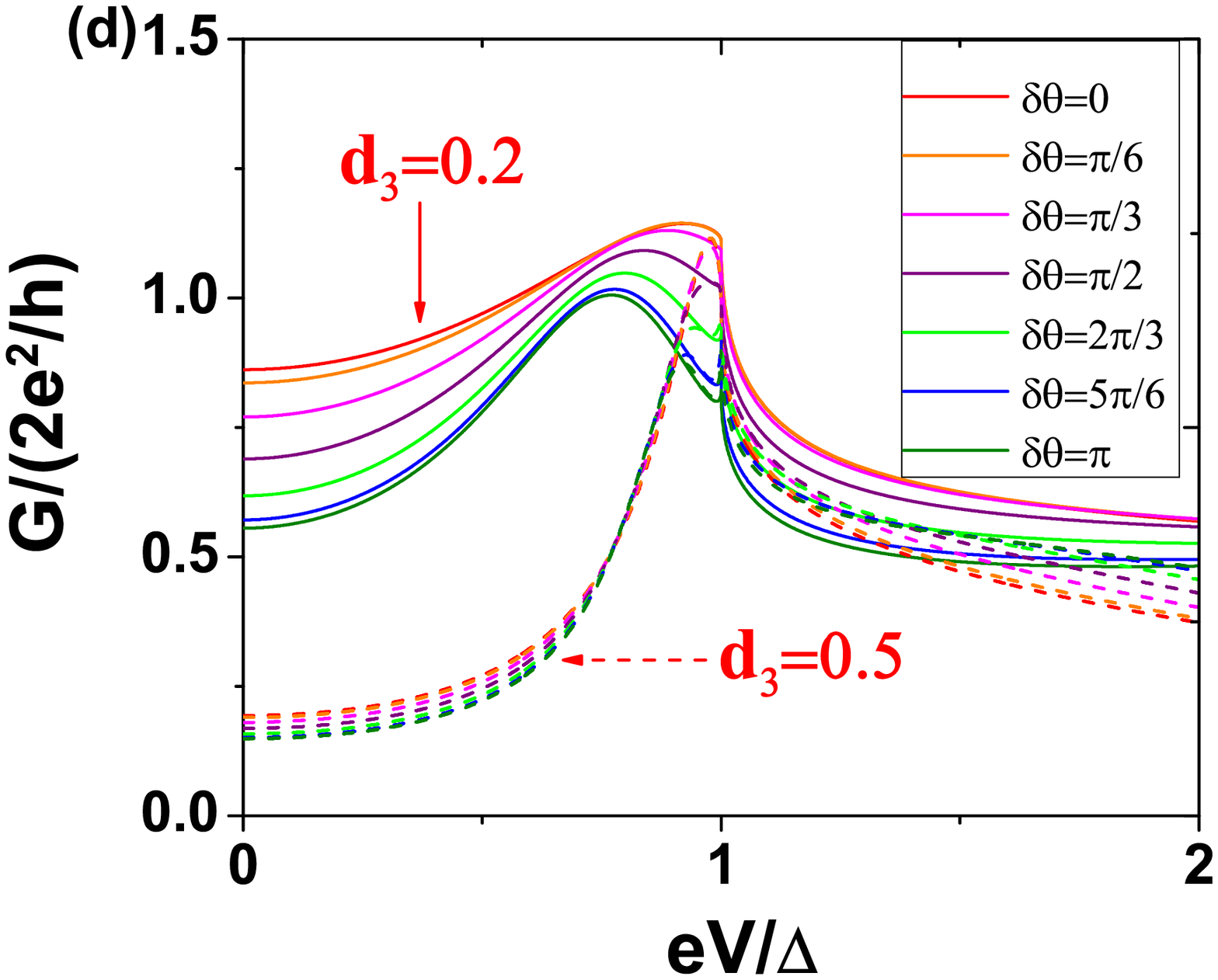}}
\caption{ Tunneling spectroscopy of FM-I-FM-I-TS junction and tunneling conductance's dependence
on the magnetization directions of FMs. Common parameters in all figures, $\mu_{f}=\mu_{I}=\mu_{s}=1$,
$\Delta_{0}=0.05$,  $M_{1}=M_{2}=0.5$, $d_{1}=0.2$, $d_{2}$=1, $Z_{1,2,3,4}=0.1$.  Tunneling spectroscopy with (a) $\theta_{1}=\theta_{2}=\pi/2$,
(b) $\theta_{1}=0$, $\theta_{2}=\pi/2$, (c) $\theta_{1}=3\pi/4$, $\theta_{2}=\pi/4$, (d) $\delta\phi=\phi_{1}-\phi_{2}=0$, $\theta_{2}=0$.
The conductance only depends on the difference of $\phi_{1}$ and $\phi_{2}$, does
not depend on the absolute value of $\phi_{1}$ and $\phi_{2}$.}
\label{fig3}
\end{figure}

\begin{figure}
\subfigure{\includegraphics[width=4cm, height=4cm]{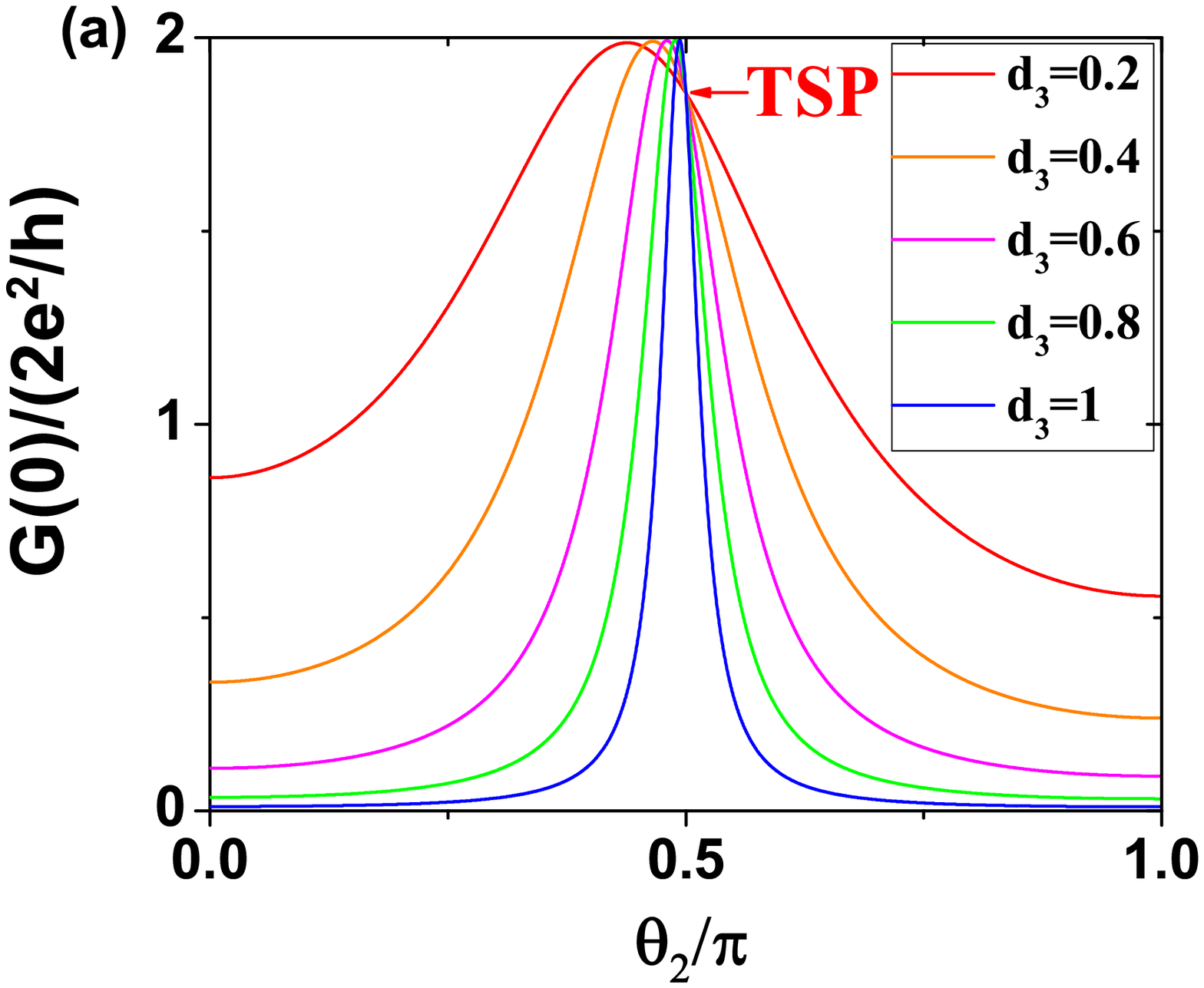}}
\subfigure{\includegraphics[width=4cm, height=4cm]{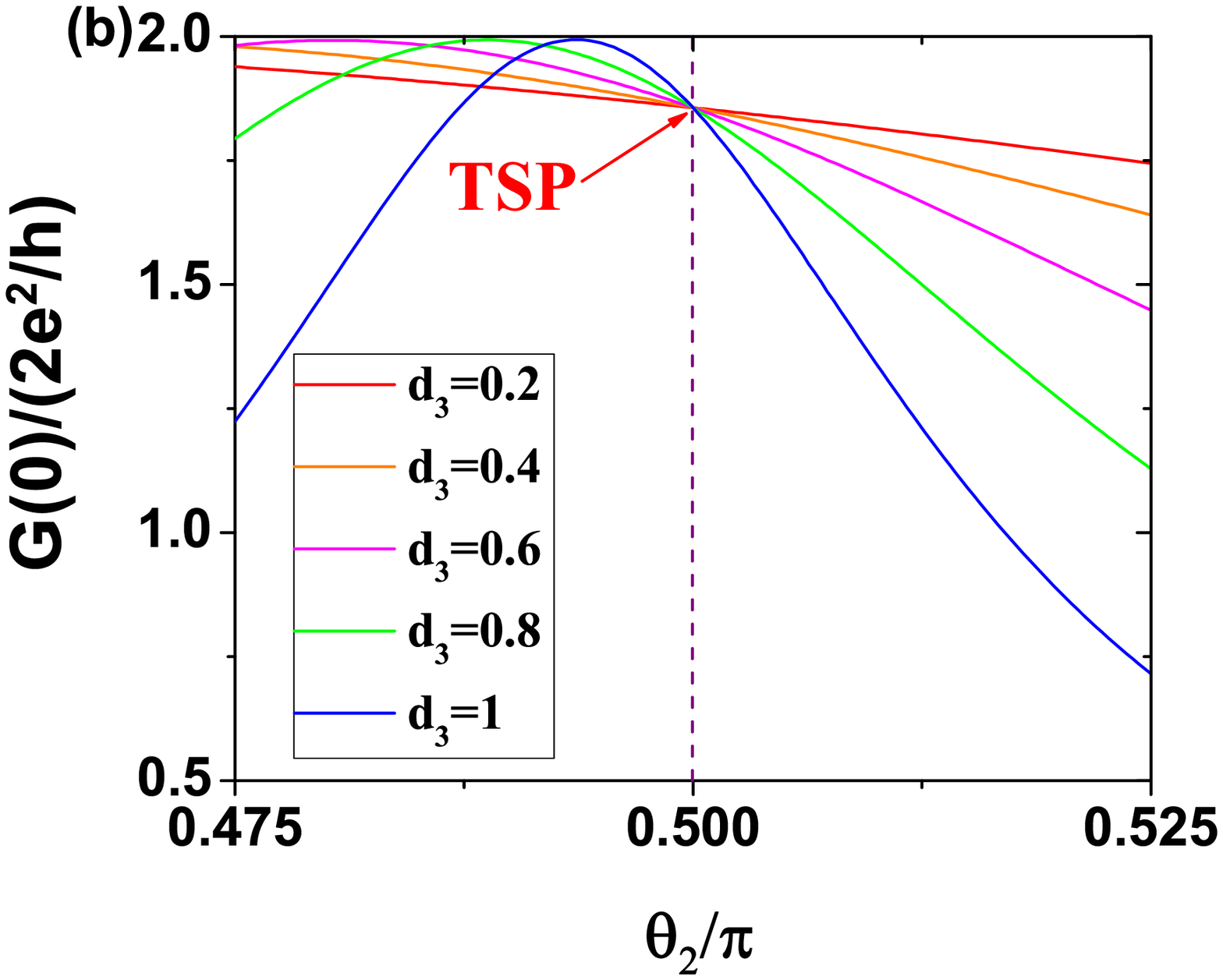}}
\subfigure{\includegraphics[width=4cm, height=4cm]{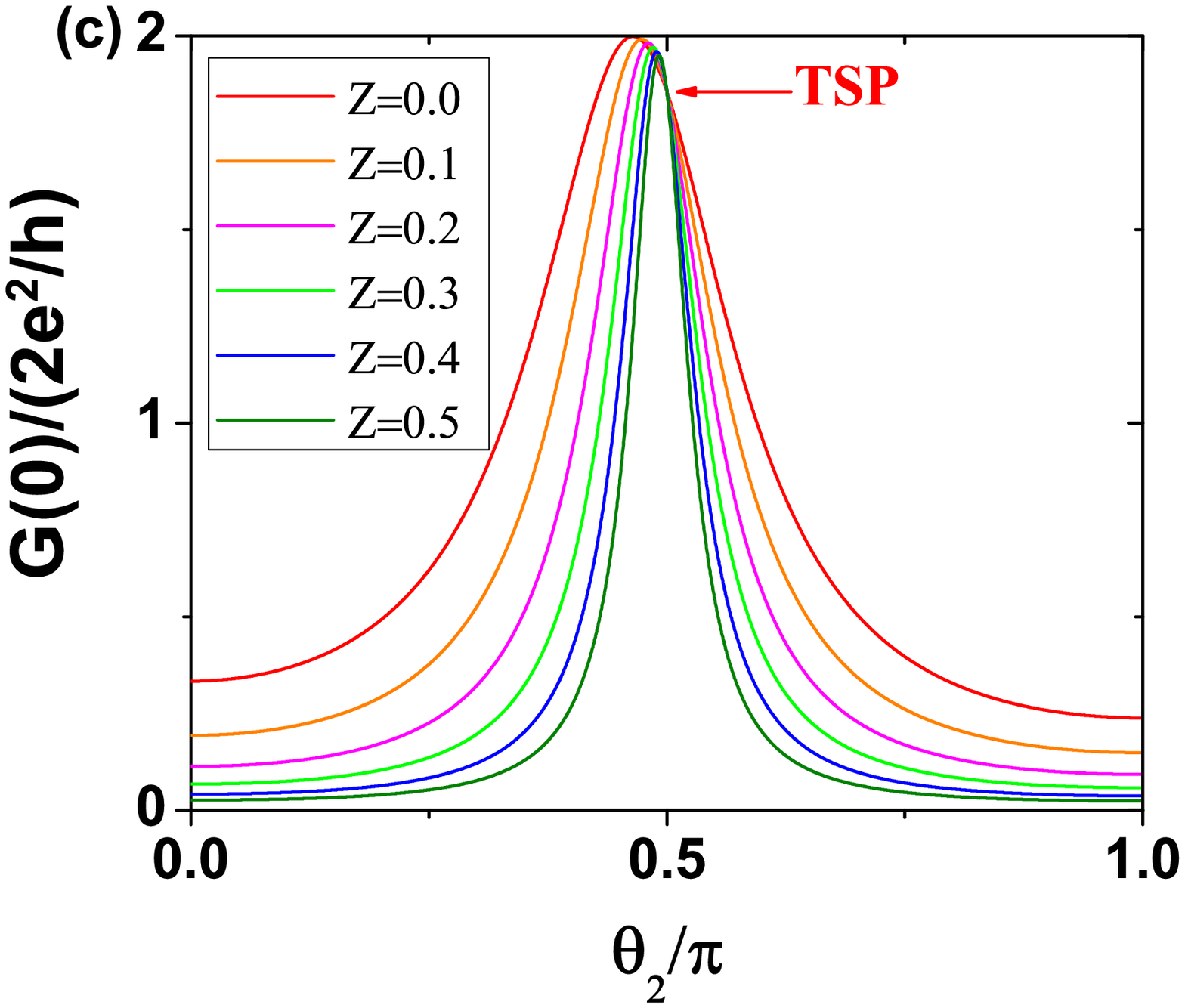}}
\subfigure{\includegraphics[width=4cm, height=4cm]{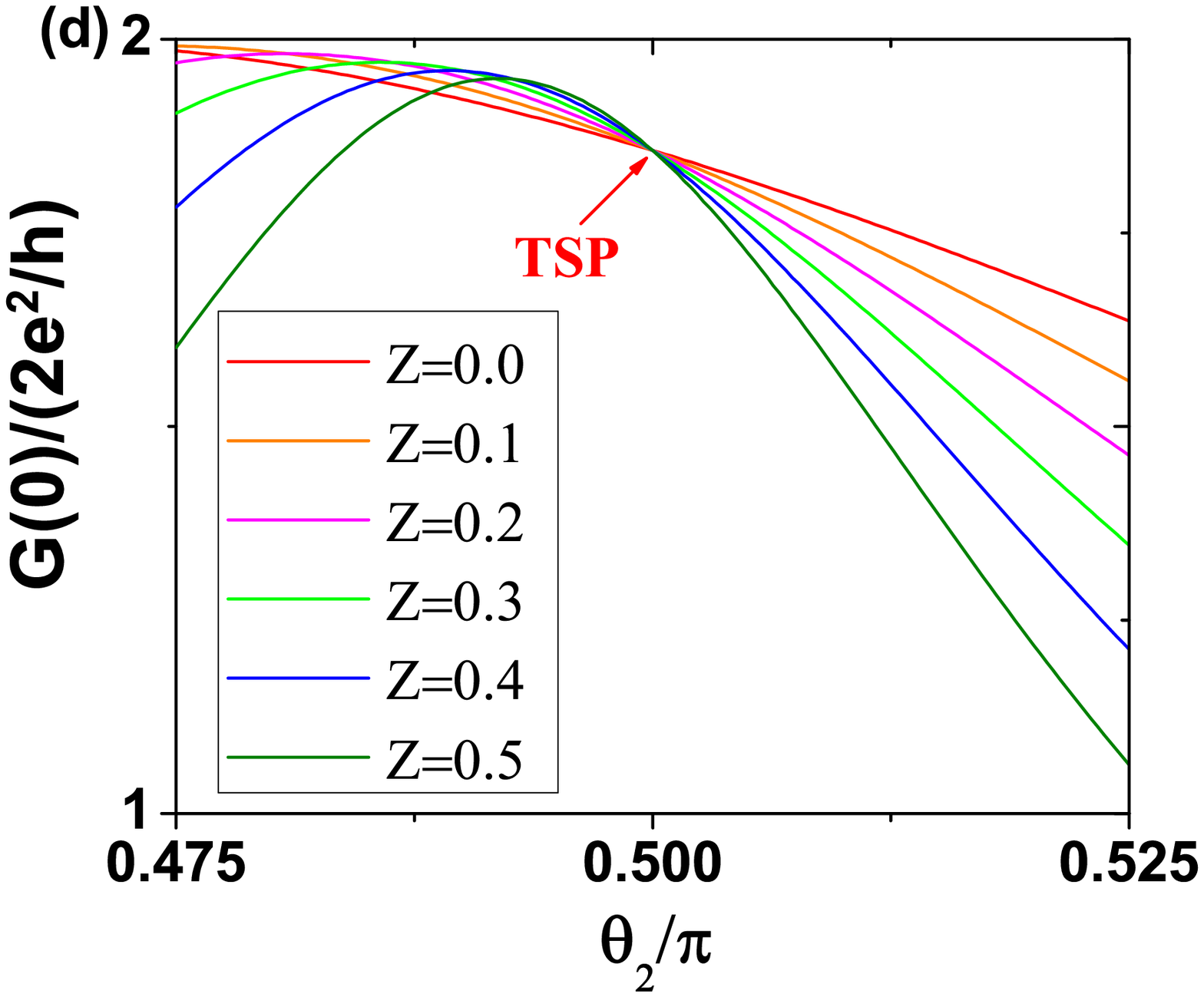}}
\caption{ZBC's dependence on the thickness of ITB1 and interface scattering potential. Common parameters in all figures, $\mu_{f}=\mu_{I}=\mu_{s}=1$,
$\Delta_{0}=0.05$,  $M_{1}=M_{2}=0.5$, $d_{2}=1$, $\theta_{1}=\phi_{1}=\phi_{2}=0$. (a)-(b) $Z_{1,2,3,4}=0.1$, $d_{1}=0.2$, (a) shows the dependence of ZBC on
the thickness of ITB1, $d_{3}$, (b) shows a closer look of the topological stable point (TSP) in (a).
(c)-(d) $d_{1}=d_{3}=0.5$, $Z_{1,2,3,4}=Z$.  (c) shows the dependence of ZBC on interface scattering potential, $Z$,
(d) shows a closer look of the TSP in (c).
}\label{fig4}
\end{figure}

The tunneling spectroscopies of the one-dimensional FM-I-FM-I-TSC junction
are shown in fig.\ref{fig3} and fig.\ref{fig4}, the most remarkable property of the tunneling
spectroscopies is that when $\theta_{2}=\pi/2$, the ZBC keeps the topological feature (as shown in
fig.\ref{fig3}(a)(b)) and is found to take the same form as eq.(\ref{9}) but with a substitution of $(k_{2+(-)},\theta_{2})$
to $(k_{1+(-)},\theta_{1})$, $i.e.$,
\begin{eqnarray}
\tilde{G}(0,\vec{n}_{1},\theta_{2}=\frac{\pi}{2})=\frac{e^{2}}{h}\frac{16k_{1+}k_{1-}}
{(k_{1+}+k_{1-})^{2}\cos^{2}\theta_{1}+4k_{1+}k_{1-}\sin^{2}\theta_{1}}\nonumber
\end{eqnarray}
with $k_{1+(-)}=\sqrt{2(\mu_{f}+(-)M_{1})}$. It is interesting
that $\tilde{G}(0,\vec{n}_{1},\theta_{2}=\pi/2)$ is independent of the magnetization strength of FM2
but only depends on the parameters of FM1, this is a manifestation of the non-local effect of the nontrivial
topology of TSC. When $\theta_{2}\neq\pi/2$, the ZBC no longer exhibits the topological feature and turns to depend
on all parameters of the whole junction.

In fig.\ref{fig3}(a), $\theta_{1}=\theta_{2}=\pi/2$, $i.e.$, $\delta\theta=\theta_{1}-\theta_{2}=0$,
by varying $\delta\phi=\phi_{1}-\phi_{2}$, it is found
that the tunneling conductance away from the zero-bias voltage shows the conventional MVE between
two FMs since the conductance in the parallel state is larger than the one in the anti-parallel state.
Similar picture also appears when $\delta\phi$ is fixed to zero while varying $\delta\theta$ with $\theta_{2}$
fixed, as shown in fig.\ref{fig3}(d). However, the results shown in fig.\ref{fig3}(b)(c) can not
be explained by the MVE between two FMs because the angle-difference between
the two magnetization direction is fixed to $\pi/2$ (the angle difference is given as
$\arccos(\cos(\delta \theta)\cos(\delta\phi))$, as $\delta \theta=\pi/2$, the varying of the
$\delta\phi$ does not change the angle-difference), this suggests that the appearance of
TSC enriches the angle-dependence. From these figures, it is also direct to see that when
$\theta_{2}\neq\pi/2$, the increase of the thickness of the ITB1, $d_{3}$, will greatly
reduce the conductance's dependence on the angle-difference between the two FMs. As we will show in
the following that the TMR grows exponentially with $d_{3}$, this
indicates that the TMR's dependence on the angle-difference between the two FMs
can be safely neglected in the high TMR regime which is of most interest.
Based on this recognition, without loss of the main physics, we set $\theta_{1}=\phi_{1}=\phi_{2}=0$,
while keeping $\theta_{2}$ as the only variable.

In fig.\ref{fig3}(c)(d), it has already shown that with the increase of
$d_{3}$, the tunneling conductance at low-bias voltage will be greatly
suppressed. Fig.\ref{fig4}(a) provides a more complete picture of
the ZBC's dependence on $\theta_{2}$ and the effect of $d_{3}$ to the ZBC.
It is clear that the minimum ZBC decreases monotonically
with $d_{3}$, but for the ZBC at $\theta_{2}=\pi/2$, the increase of $d_{3}$
has no effect to it, as shown in Fig.\ref{fig4}(b). From the expression of
$\tilde{G}(0,\vec{n}_{1},\theta_{2}=\pi/2)$ we know it takes the value
$16k_{1+}k_{1-}G_{0}/(k_{1+}+k_{1-})^{2}$.
Fig.\ref{fig4}(c)(d) show the effect of the interface scattering potential to
the ZBC, it is easy to see that the results are similar to fig.\ref{fig4}(a)(b).

\begin{figure}[t]
\subfigure{\includegraphics[width=4cm, height=4cm]{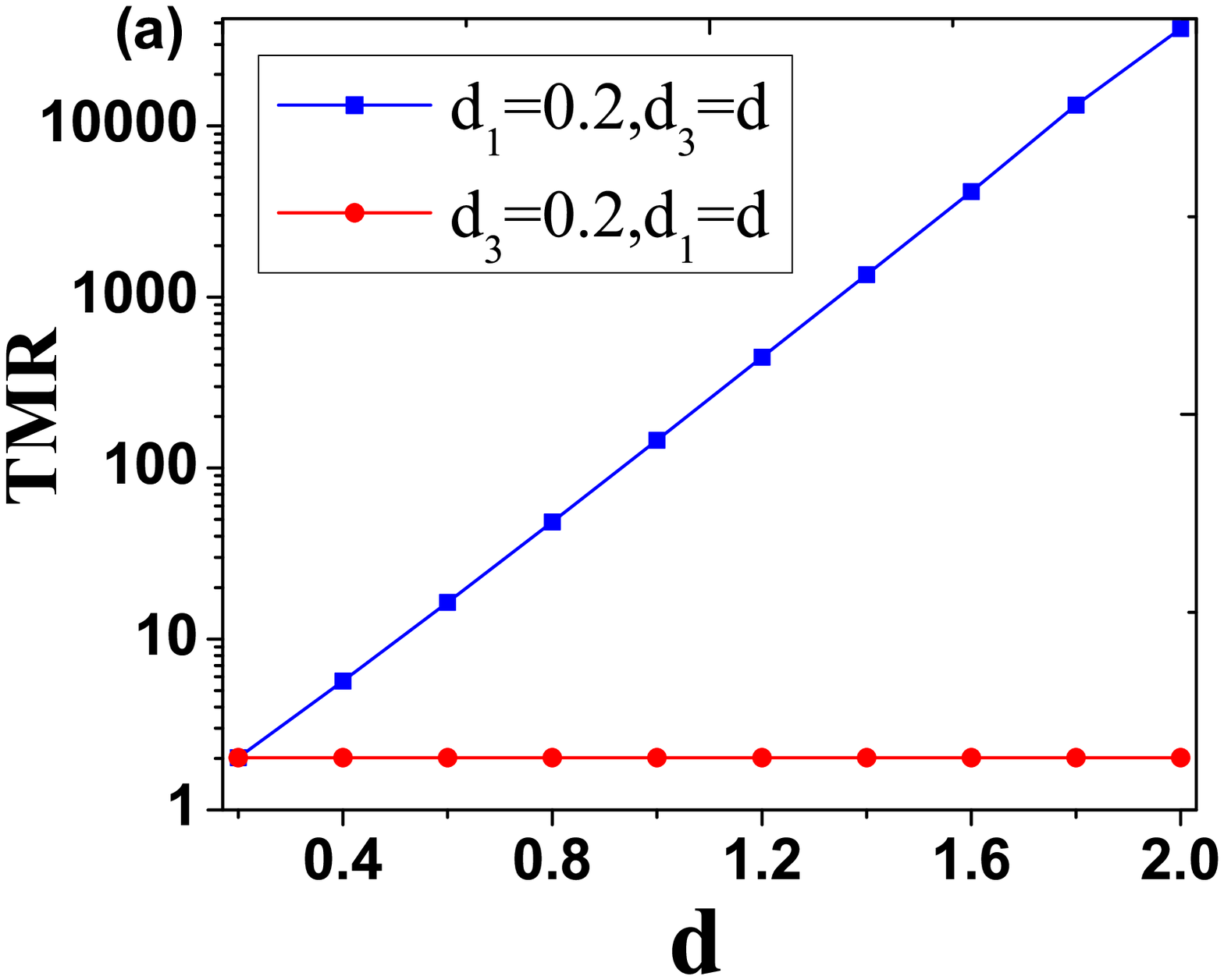}}
\subfigure{\includegraphics[width=4cm, height=4cm]{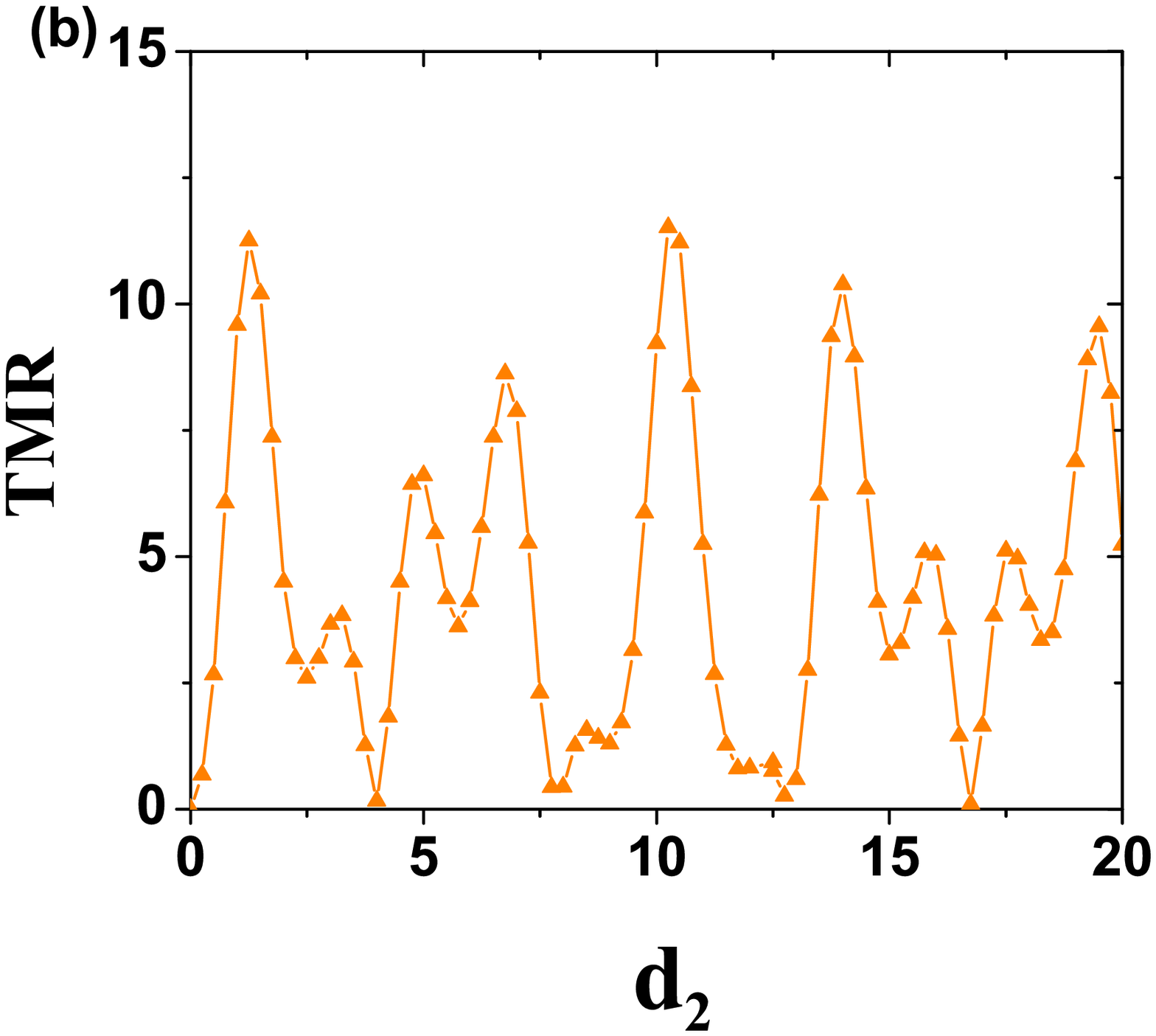}}
\subfigure{\includegraphics[width=4cm, height=4cm]{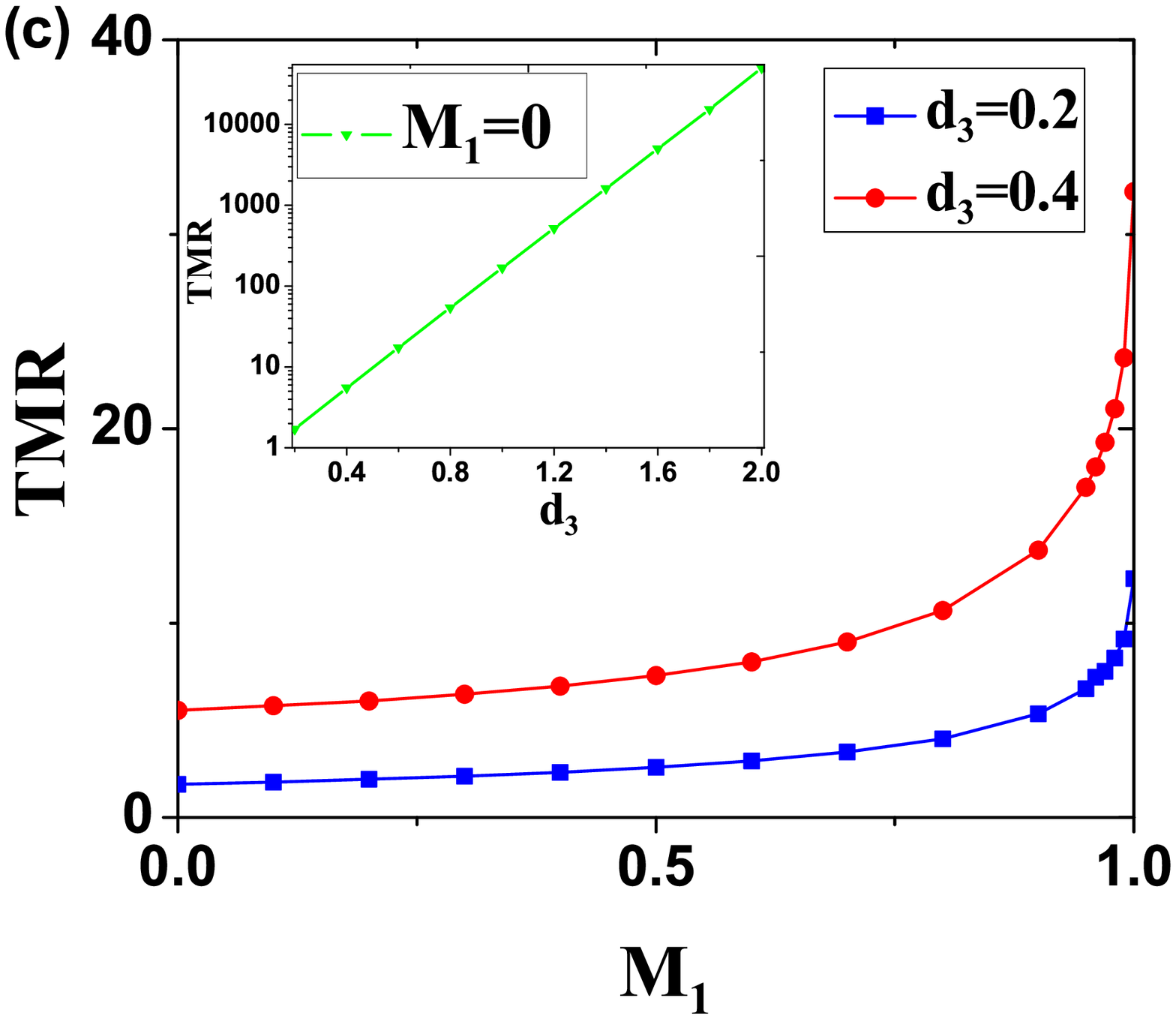}}
\subfigure{\includegraphics[width=4cm, height=4cm]{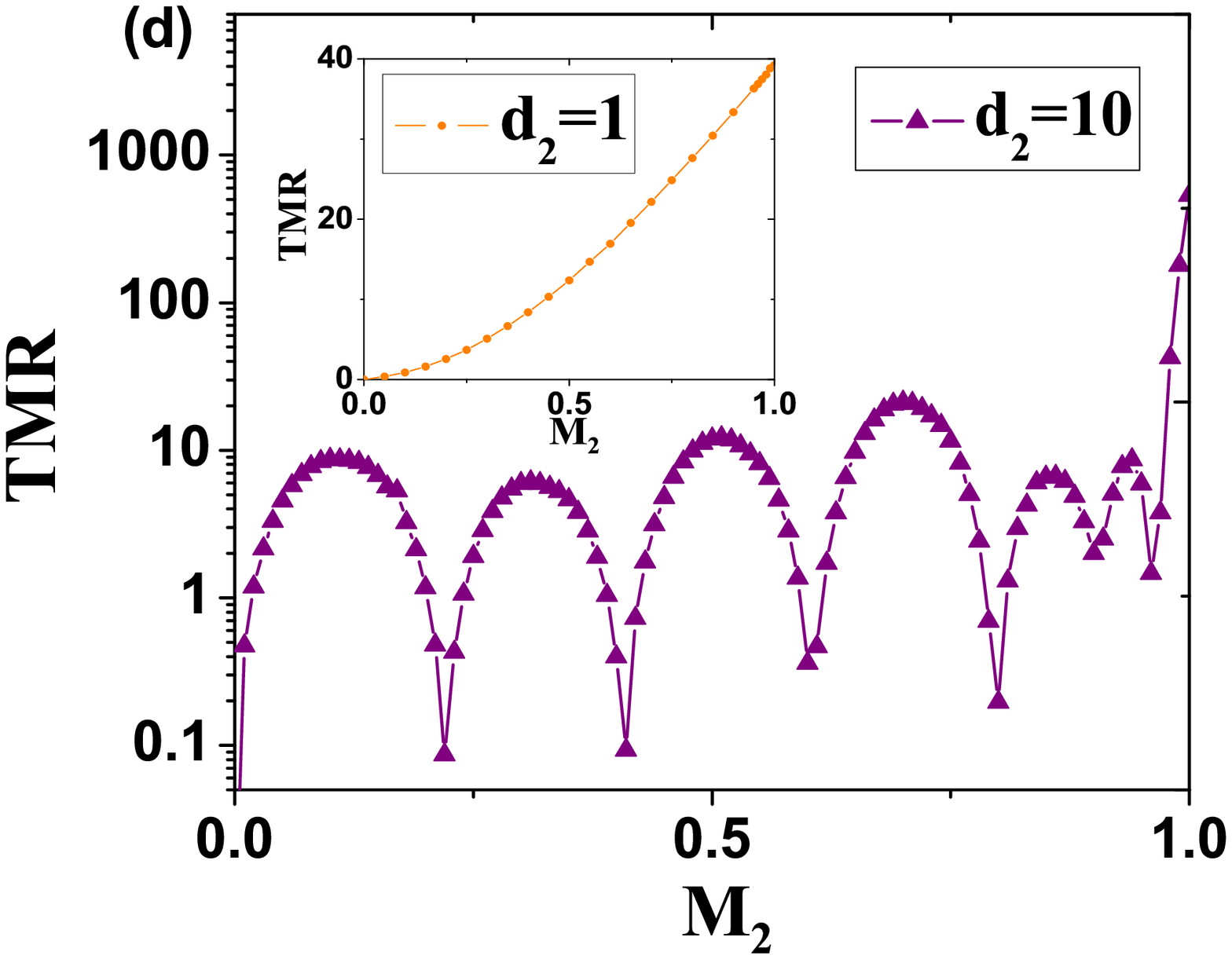}}
\caption{ TMR's dependence on parameters of FM-I-FM-I-TS junction. Common parameters in all figures, $\mu_{f}=\mu_{I}=\mu_{s}=1$,
$\Delta_{0}=0.05$, $\theta_{1}=\phi_{1}=\phi_{2}=0$. (a) TMR's dependence on the thickness of ITB, $M_{1}=M_{2}=0.5$, $d_{2}=1$. (b) TMR's dependence on the length of FM2, $M_{1}=M_{2}=0.5$, $d_{1}=d_{3}=0.5$. (c) TMR's dependence on the magnetization strength of FM1, $M_{2}=0.5$,  $d_{1}=0.2$, $d_{2}=1$.
(d) TMR's dependence on the magnetization strength of FM2, $M_{1}=0.5$,  $d_{1}=d_{3}=0.5$. In (a)(d), logarithmic 
coordinates are used.}\label{fig5}
\end{figure}

Since the increase of $d_{3}$ and the increase of interface scattering potential both
have no effect to the ZBC at $\theta_{2}=\pi/2$, but
will greatly suppress the ZBC away from $\theta_{2}=\pi/2$, it is obvious that
the increase of $d_{3}$ and the increase of interface scattering potential will greatly
increase the TMR. In fig.\ref{fig5}(a), the blue square-line shows that the TMR
grows exponentially with the increase of $d_{3}$. This is a remarkable property,
because it indicates that the TMR is easy to tune and can be tuned to arbitrary
high value even the two FMs are both weak. Unlike the conventional MVE
between two FMs that the increase
of the thickness of ITB will exponentially suppress the tunneling conductance (or
say tunneling current) for all possible angle-difference, here due to the nontrivial
topological property of the TSC, the tunneling conductance at the neighbourhood of  $\theta_{2}=\pi/2$ is
topologically stable and therefore, even $d_{3}$ is large, the tunneling current
can still be large. This is also important for real applications because the strength of
the tunneling current is proportional to the strength of signal, therefore a high signal-to-noise
ratio, which is necessary for storage applications~\cite{Chappert:2007}, can be guaranteed in this system.
The TMR's explicit dependence on
the interface scattering potential of each interface is neglected here, the only thing we
need to stress again is that the TMR monotonically increases with the increase of
interface scattering potential, and therefore, the interface roughness  emerging in the process of
synthetizing the junction is not bad here.

Other parameters of the junction also may affect the TMR. The red dot-line in fig.\ref{fig5}(a)
shows the TMR's dependence on the thickness of ITB1, $d_{1}$, we can see that the TMR is almost unchanged
with the variation of $d_{1}$. Although an increase of $d_{1}$ has small impact on the TMR,
it can greatly suppress the magnetic proximity effect of FM2 to the TSC. Fig.\ref{fig5}(b) shows that
the TMR has an oscillating dependence on
the length of the FM2, $d_{2}$, this oscillating behavior is due to interference of wave functions in
FM2. In fig.\ref{fig5}(c), the TMR shows a dependence on $M_{1}$ which is similar to the inset of fig.\ref{fig1}(d),
where the TMR monotonically increases with the increase of the magnetization
strength. The inset of fig.\ref{fig5}(c) shows a remarkable property of the junction that even
the FM1 turns to be a normal metal (NM), the TMR also grows exponentially with the increase of $d_{3}$.
Therefore, a NM-I-FM-I-TSC junction is also an ideal structure for applications.
In fig.\ref{fig5}(d), it is shown that when the length of FM2 is large, $i.e.$,
$k_{2+(-)}d_{2}>>\pi$, TMR will exhibit an oscillating dependence on $M_{2}$,
which is also due to interference of wave functions in FM2. However, when $k_{2+}d_{2}<\pi$,
TMR will exhibit a monotonically increasing dependence on $M_{2}$, as shown in the inset of
fig.\ref{fig5}(d). From fig.\ref{fig5}(b)(d), we see that  the interference of wave functions in
FM2 may greatly suppressed the TMR,  to avoid the suppression in real applications, therefore,
the length of FM2 is better to choose to satisfy $k_{2+}d_{2}<\pi$.

\section{V. Higher-dimensional case}
\label{V}

Here we only consider a two-dimensional
FM-I-FM-I-TSC junction for illustration, the three-dimensional case can be
directly generalized from it.

The two-dimensional Hamiltonian  we consider under
the representation $\hat{\Psi}^{\dag}(\vec{r})
=(\hat{\psi}^{\dag}_{\uparrow}(\vec{r}),\hat{\psi}_{\downarrow}(\vec{r}),\hat{\psi}^{\dag}_{\downarrow}(\vec{r}),
\hat{\psi}_{\uparrow}(\vec{r}))$ is given as
\begin{eqnarray}
\mathcal{H}_{2D}=\tau_{z}\left[-\frac{\partial^{2}_{x}+\partial^{2}_{y}}{2} - \mu(x) + V(x)
\right] + \tau_{x}\Delta(x)+\tau_{y}\Delta(y), \nonumber
\end{eqnarray}
here we have assumed that the chemical potential $\mu(\vec{r})$, the potential $V(\vec{r})$
and the pairing potential $\Delta(x)$
take the same form as $\mu(x)$, $V(x)$ and $\Delta(x)$ given previously, respectively.
The other pairing potential $\Delta(y)=i\Delta_{0}\Theta(x-x_{1})\partial_{y}$.

When the dimension of the system is higher than one, the injected electron's momentum
$\vec{k}$ can be decomposed as
$(k_{x}, k_{\perp})$, for the two-dimensional case we consider here, $k_{\perp}=k_{y}$.
If $k_{\perp}$ is conserved in the process of electrons transporting across the junction,
the tunneling process of an injected electron with nonzero $k_{\perp}$ at zero-bias voltage
no longer exhibits the topological feature. This can be simply explained under the Majorana picture~\cite{Law:2009,Flensberg:2010}.
In higher dimension, the boundary Majorana zero modes require
$k_{\perp}=0$, this indicates that due to the momentum conservation, an injected electron with
nonzero $k_{\perp}$ will no longer directly couple with the Majorana zero modes, and then the ZBC will
be parameter-dependent (topological feature is absent) no matter what value $\theta_{2}$ takes. As the
final conductance includes contributions from every direction, it is obvious that the new added dimension
will generally mask the topological effect. However, as the conductance curve should be continuous, this
implies that if the ratio of the transverse momentum to the longitudinal momentum in TSC decreases, the effect of
the non-zero transverse momentum will decrease, and then the topological effect will be strengthened.
One way to reduce the momentum ratio in TSC is to increase $\mu_{s}$, since $k_{s,\perp}=k_{\perp}$ is independent of
$\mu_{s}$, while $|k_{s,x}|$ monotonically increases with $\mu_{s}$.

\begin{figure}[t]
\subfigure{\includegraphics[width=4cm, height=4cm]{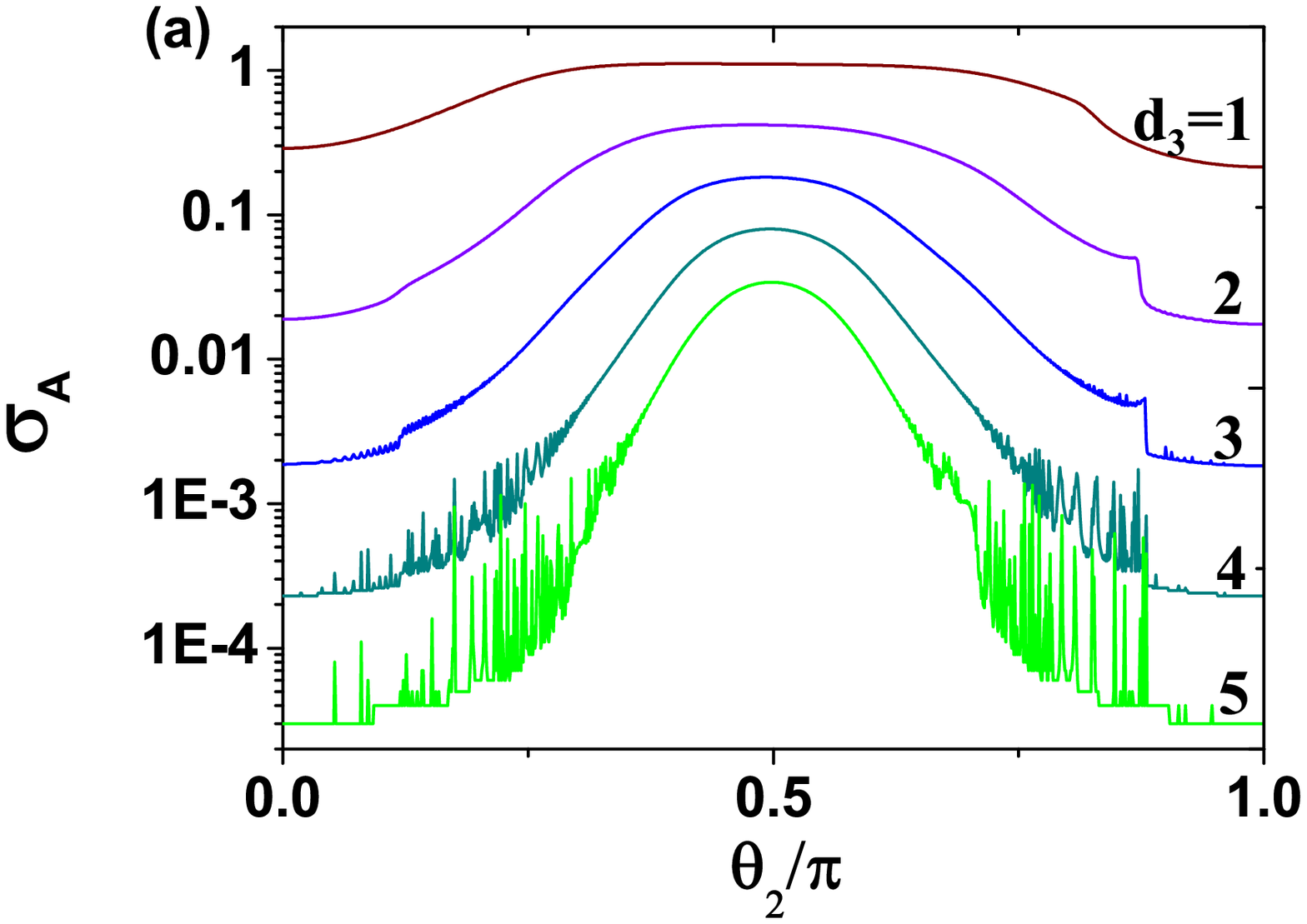}}
\subfigure{\includegraphics[width=4cm, height=4cm]{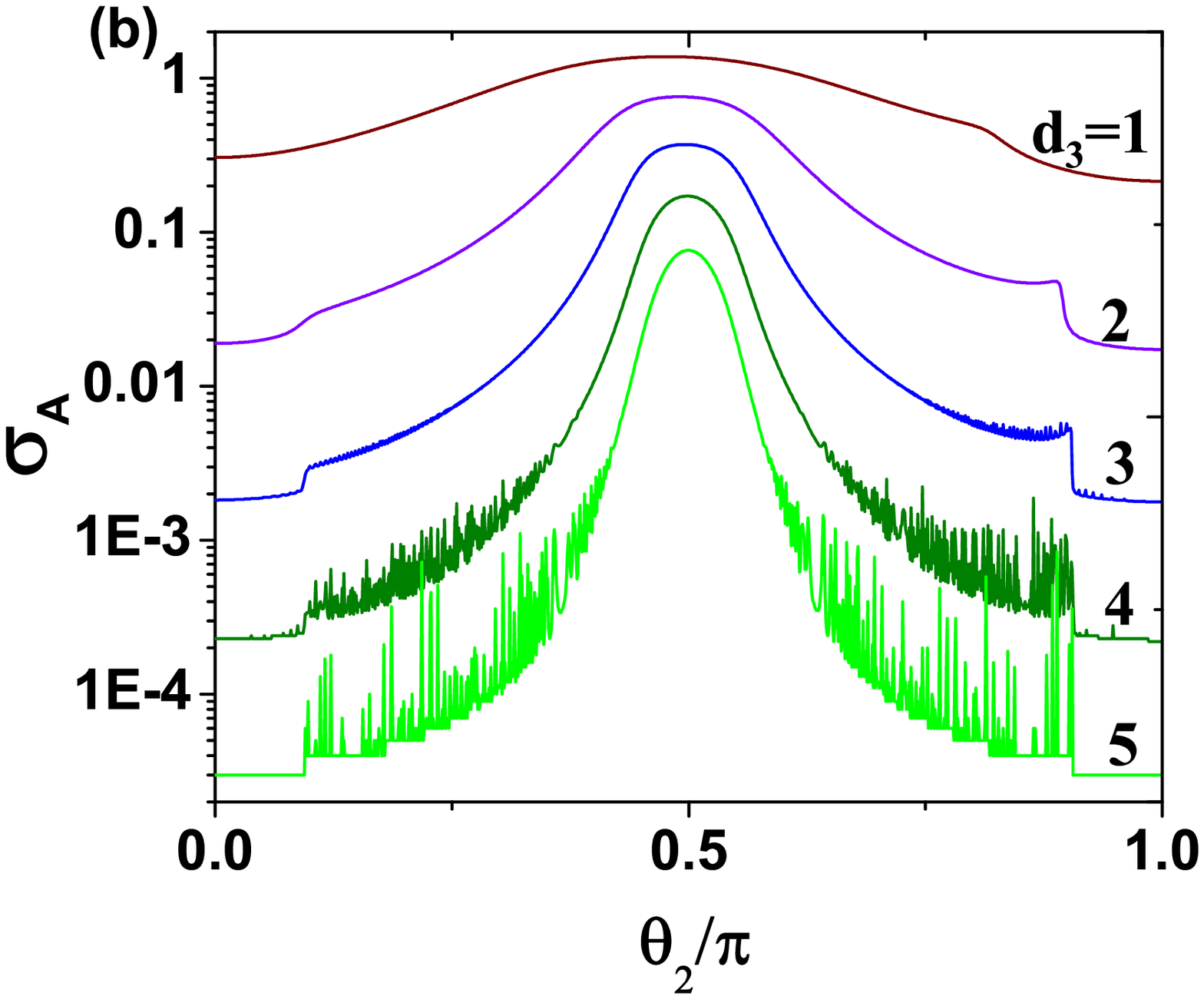}}
\subfigure{\includegraphics[width=4cm, height=4cm]{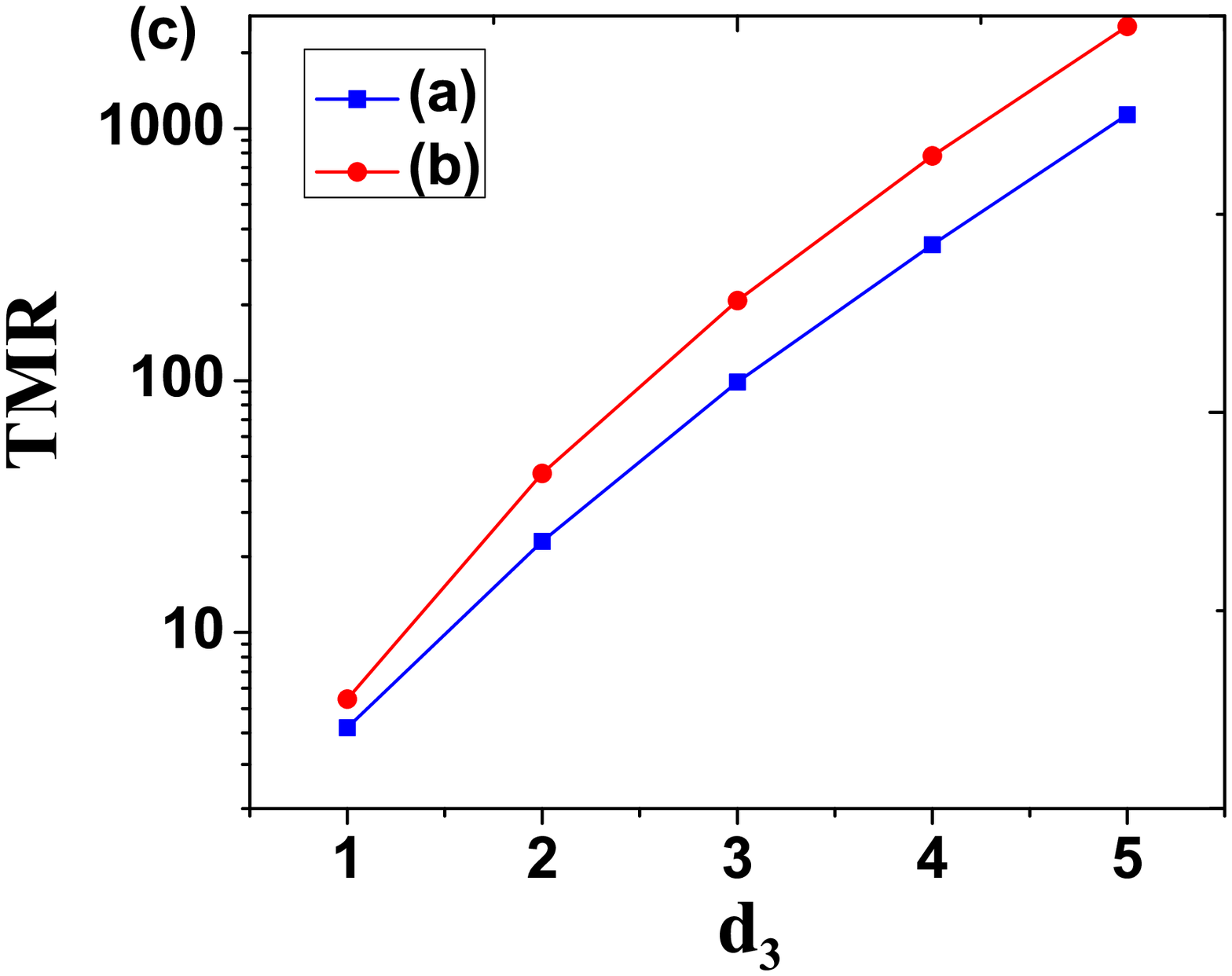}}
\subfigure{\includegraphics[width=4cm, height=4cm]{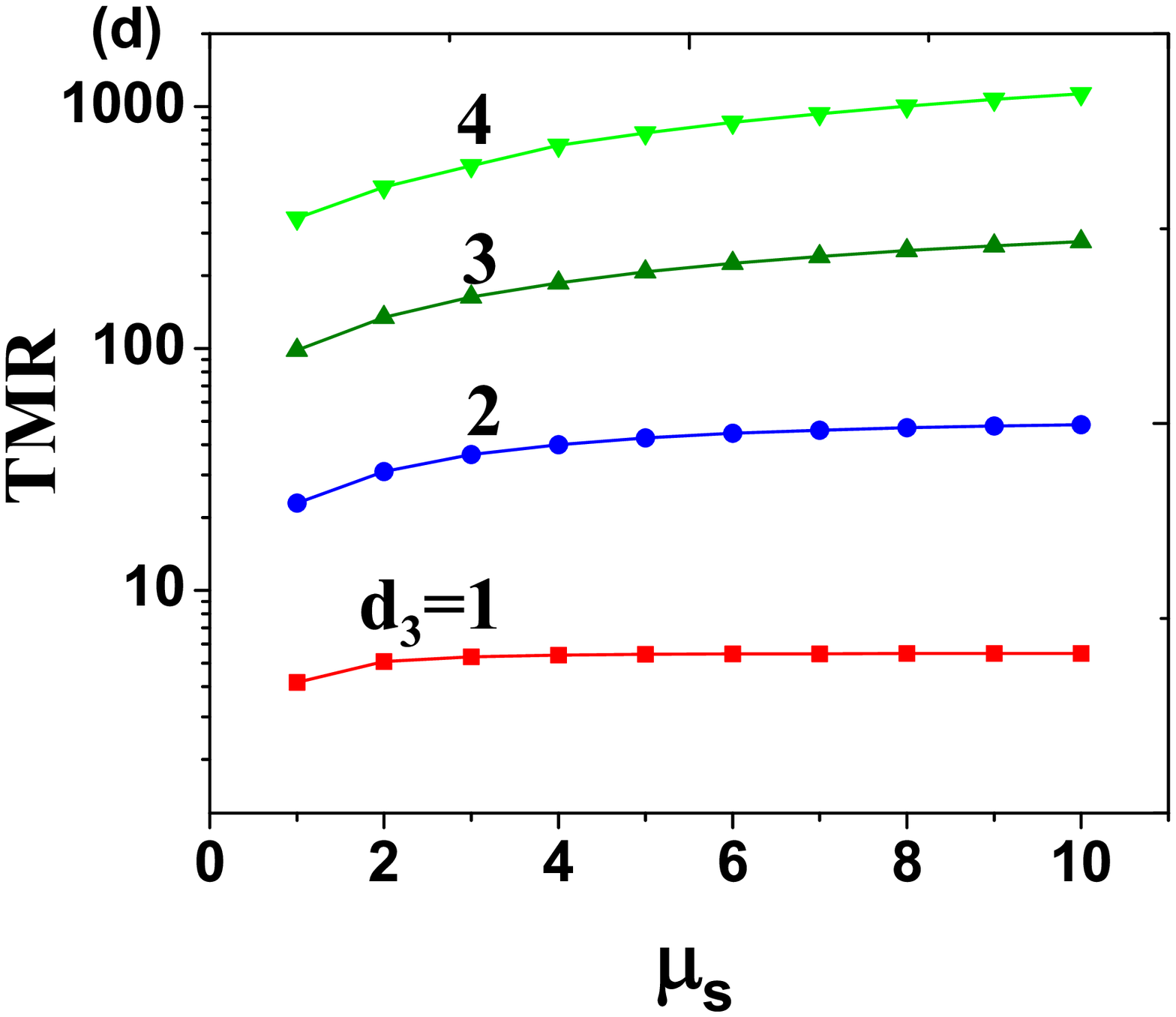}}
\caption{Tunneling spectroscopy of FM-I-FM-I-TS junction in two dimension and TMR's dependence on
parameters, logarithmic coordinates are used in all figures. Common parameters in all figures, $\mu_{f}=1$, $\mu_{I}=0.1$,
$\Delta_{0}=0.05$, $M_{1}=M_{2}=0.4$, $d_{1}=0.2$, $d_{2}=1$, $\theta_{1}=\phi_{1}=\phi_{2}=0$. Here choosing a small $\mu_{I}$
is to enhance the momentum filtering effect. (a)-(b) show the ZBC's dependence on $\theta_{2}$ and $d_{3}$ with (a) $\mu_{s}=1$, (b) $\mu_{s}=5$. (c)
shows the TMR-$d_{3}$ relation corresponding to (a) and (b). (d)
shows the TMR-$\mu_{s}$ relation for different $d_{3}$.
 }\label{fig6}
\end{figure}

The junction itself has an intrinsic effect to enhance the topological effect.
The intrinsic effect is the so-called momentum filtering effect (MFE)~\cite{Yuasa:2000} that is if $\mid\vec{k}\mid$ is fixed, then injected electrons with larger $k_{x}$ will feel a lower
tunneling barrier because the effective
potential barrier is given as $U_{I,eff}=\mu_{I}+|\vec{k}|^{2}-k_{x}^{2}=\mu_{I}+|\vec{k}|^{2}\sin^{2}\theta$ with
$\theta=\arccos(k_{x}/|\vec{k}|)$. As the MFE reduces
the contributions from injected electrons with larger $k_{\perp}$, the ratio of the contribution from
the neighbourhood of  $k_{\perp}=0$ increases, consequently, the topological effect is also enhanced. Therefore,
when the MFE is very strong, the TMR is expected to be very large. From the
expression of $U_{I,eff}$, we know that the most direct way to enhance the MFE is to reduce $\mu_{I}$,
this can be realized by choosing small gap insulating materials as the ITB.

For the sake of discussion, we introduce a dimensionless quantity $\sigma_{A}$ which is defined as
\begin{eqnarray}
\sigma_{A}(\theta_{2})=\frac{\sum_{k_{y}}\bar{G}(0,\theta_{2},k_{y})}{\sum_{k_{y}}2e^{2}/h}
\end{eqnarray}
where $\bar{G}(0,\theta_{2},k_{y})$ is the summation of ZBC for spin-up and spin-down electron with transverse momentum $k_{y}$.
Fig.\ref{fig6}(a)(b) shows that with the increase of $d_{3}$, ZBC will be greatly suppressed no matter what value $\theta_{2}$ takes,
this agrees with our previous argument. However, due to a combination of the topological effect and MFE, the suppression at the
neighbourhood of $\theta_{2}=\pi/2$ is still much smaller than the ones at the neighbourhood of $\theta_{2}=0$ or $\pi$.
The angle-dependent suppression behavior results in an approximately exponential dependence between the TMR (now given as
$(\sigma_\mathrm{A,max}-\sigma_\mathrm{A,min})/\sigma_\mathrm{A,min}$) and $d_{3}$, as
shown in fig.\ref{fig6}(c), therefore, the remarkable property of the FM-I-FM-I-TSC junction that the TMR
is easy to tune and can be tuned to arbitrary high value is still hold in two-dimension.
Compared fig.\ref{fig6}(b) to fig.\ref{fig6}(a), we can see that an increase of $\mu_{s}$ indeed results
in an enhancement of the topological effect and consequently a slower decrease of $\sigma_{A}$ at the neighbourhood of $\theta_{2}=\pi/2$.
Fig.\ref{fig6}(c)(d) show that the slower decrease of $\sigma_{A}$ results in a larger TMR.  The
effects of other parameters are similar to the ones shown in fig.\ref{fig5}, we do not
plan to discuss them here again.

\section{VI. Discussions and conclusions}
\label{VI}

In summary,  because the tunneling process in junctions
composed of FM and time-reversal invariant TSC without SRS strongly depends on
the spin-polarization direction of the injected electrons, high TMR is found to exist
in these junctions. Compared to  conventional TMR shown in MTJs, the TMR shown in FM-I-TSC junction and
FM-I-FM-I-TSC junction exhibits several extraordinary
characteristics: (i) so far, the best mechanism to obtain high TMR in
conventional MTJs is to take use of a MgO tunneling barrier~\cite{Butler:2001, Mathon:2001, Parkin:2004, Yuasa:2004}, however, for junctions
we have considered,  high TMR is obtained even the
ITB is featureless;  (ii) for the FM-I-TSC junction in one dimension, the
TMR only depends on the magnetization strength of the FM, and it goes to infinity when the
FM turns to be a half metal; (iii) for the FM-I-FM-I-TSC junction, the TMR shows a remarkable property that it grows exponentially with
the thickness of ITB between the two FMs and also grows  monotonically with the interface scattering potential,
this remarkable property makes it possible to tune the
TMR to arbitrary high value even the magnetization strength of the two FMs are both weak, in fact,
the magnetization strength of FM1 can even be zero.
Even with a consideration of the effect of finite temperature, these characteristics will still hold if
$k_{B}T$ is much smaller than the energy gap of the TSC.

In short, a combination of the FM and TSC provides a new fascinating mechanism to obtain high TMR in a
convenient way. Besides this remarkable effect, we also note there already exist some works pointing
out that spin-polarized current, another important element for spintronics, can also be simply realized by
making use of TSCs\cite{He:2014, Tanaka:09}. All these results suggest that the nontrivial topological property
of TSCs may bring new insights in spintronics.

\section{VII. Acknowledgements}

We thank Zhong Wang for useful discussions. This work is supported by NSFC under Grant
 No.11275180.

\section{Appendix}

When a spin-up electron with energy $E$ is injected from the FM1, the wave functions
in the two ferromagnetic parts are given as
\begin{eqnarray}
&&\psi_{FM1}=\vec{\chi}_{1}e^{ik_{1+,e}x}+b_{1+}\vec{\chi}_{1}e^{-ik_{1+,e}x}+
a_{1+}\vec{\chi}_{2}e^{ik_{1+,h}x}\nonumber\\
&&\qquad\quad+b_{1-}\vec{\chi}_{3}e^{-ik_{1-,e}x}
+a_{1-}\vec{\chi}_{4}e^{ik_{1-,h}x},\nonumber\\
&&\psi_{FM2}=b_{3L+}\vec{\chi}_{1}^{'}e^{-ik_{2+,e}x}+b_{3R+}\vec{\chi}_{1}^{'}e^{ik_{2+,e}x}
+a_{3L+}\vec{\chi}_{2}^{'}e^{-ik_{2+,h}x}\nonumber\\
&&\qquad\quad+a_{3R+}\vec{\chi}_{2}^{'}e^{ik_{2+,h}x}+b_{3L-}\vec{\chi}_{3}^{'}e^{-ik_{2-,e}x}+
b_{3R-}\vec{\chi}_{3}^{'}e^{ik_{2-,e}x}\nonumber\\
&&\qquad\quad+a_{3L-}\vec{\chi}_{4}^{'}e^{-ik_{2-,h}x}
+a_{3R-}\vec{\chi}_{4}^{'}e^{ik_{2-,h}x}, \nonumber
\end{eqnarray}
with $\vec{\chi}_{1}=(\eta_{1},0,\eta_{2}e^{i\phi_{1}},0)^{T}$, $\vec{\chi}_{2}=(0,\eta_{2},0,\eta_{1}e^{i\phi_{1}})^{T}$,
$\vec{\chi}_{3}=(-\eta_{2},0,\eta_{1}e^{i\phi_{1}},0)^{T}$, $\vec{\chi}_{4}=(0,-\eta_{1},0,\eta_{2}e^{i\phi_{1}})^{T}$,
where $\eta_{1}=\cos(\theta_{1}/2)$, $\eta_{2}=\sin(\theta_{1}/2)$;
$\vec{\chi}_{i}^{'}=\vec{\chi}_{i}(\theta_{1},\phi_{1}\rightarrow\theta_{2},\phi_{2})$;
the momenta $k_{1,2+,e(h)}=\sqrt{2(\mu_{f}+M_{1,2}+(-)E)}$, $k_{1,2-,e(h)}=\sqrt{2(\mu_{f}-M_{1,2}+(-)E)}$.
The wave functions
in the two insulating parts are given as
\begin{eqnarray}
&&\psi_{I1}=b_{2L\uparrow}\vec{e}_{1}e^{-k_{I,e}x}+b_{2R\uparrow}\vec{e}_{1}e^{k_{I,e}x}
+a_{2L\downarrow}\vec{e}_{2}e^{-k_{I,h}x}\nonumber\\
&&\qquad\quad+a_{2R\downarrow}\vec{e}_{2}e^{k_{I,h}x}+b_{2L\downarrow}\vec{e}_{3}e^{-k_{I,e}x}+
b_{2R\downarrow}\vec{e}_{3}e^{k_{I,e}x}\nonumber\\
&&\qquad\quad+a_{2L\uparrow}\vec{e}_{4}e^{-k_{I,h}x}
+a_{2R\uparrow}\vec{e}_{4}e^{k_{I,h}x},\nonumber\\
&&\psi_{I2}=\psi_{I1}(b_{2\alpha\beta},a_{2\alpha\beta}\rightarrow b_{4\alpha\beta},a_{4\alpha\beta}),\alpha=R,L;\beta=\uparrow,\downarrow,\nonumber
\end{eqnarray}
where $\vec{e}_{1}=(1,0,0,0)^{T}$, $\vec{e}_{2}=(0,1,0,0)^{T}$, $\vec{e}_{3}=(0,0,1,0)^{T}$,
$\vec{e}_{4}=(0,0,0,1)^{T}$, the momenta $k_{I,e(h)}=\sqrt{2(\mu_{I}-(+)E)}$. The wave functions
in the superconducting part is given as
\begin{eqnarray}
&&\psi_{TS}=t_{5e}\vec{\lambda}_{1}e^{ik_{s,e}x}
+t_{5h}\vec{\lambda}_{2}e^{ik_{s,h}x}+s_{5e}\vec{\lambda}_{3}e^{ik_{s,e}x}\nonumber\\
&&\qquad\quad+s_{5h}\vec{\lambda}_{4}e^{ik_{s,h}x}\nonumber
\end{eqnarray}
with $\vec{\lambda}_{1}=(u_{1k},v_{1k},0,0)^{T}$, $\vec{\lambda}_{2}=(u_{2k},v_{2k},0,0)^{T}$,
$\vec{\lambda}_{3}=(0,0,u_{1k},v_{1k})^{T}$, $\vec{\lambda}_{4}=(0,0,u_{2k},v_{2k})^{T}$,
where $u_{1(2),k}=\Delta_{0} k_{s,e(h)}/N_{1(2)}$, $v_{1(2),k}=(E+\mu_{s}-k_{s,e(h)}^{2}/2)/N_{1(2)}$
with $N_{1(2)}=\sqrt{|u_{1(2),k}|^{2}+|v_{1(2),k}|^{2}}$ the normalization coefficients.
$k_{s,e(h)}=(-)\sqrt{Q_{1}+Q_{2}}+i\sqrt{Q_{1}-Q_{2}}$
with $Q_{1}=(\mu_{s}-\Delta^{2})$ and $Q_{2}=\sqrt{\mu_{s}^{2}-E^{2}}$ (when $E<\Delta\sqrt{2\mu_{s}-\Delta^{2}}$),
or $k_{s,e(h)}=(-)\sqrt{2(Q_{1}+Q_{3})}$ with $Q_{3}=\sqrt{E^{2}+\Delta^{4}-2\mu_{s}\Delta^{2}}$ (when $E>\Delta\sqrt{2\mu_{s}-\Delta^{2}}$).


\end{document}